\documentclass[fleqn,usenatbib]{mnras}

\usepackage{amssymb}

\usepackage{newtxtext,newtxmath}
\usepackage{accents}
\newcommand{\ubar}[1]{\underaccent{\bar}{#1}}

\usepackage[T1]{fontenc}

\DeclareRobustCommand{\VAN}[3]{#2}
\let\VANthebibliography\thebibliography
\def\thebibliography{\DeclareRobustCommand{\VAN}[3]{##3}\VANthebibliography}

\usepackage{graphicx}	
\usepackage{amsmath}	
\usepackage{amssymb}	
\newcommand\rote{\rotatebox[origin=c]{180}{$\mathrm{e}$}}
\usepackage{tipa}

\title[X-ray AGNs in SRG/eROSITA galaxy mergers]{X-ray AGNs with SRG/eROSITA: Multi-wavelength observations reveal merger triggering and post-coalescence circumnuclear blowout}
\author[R. W. Bickley et al.]{Robert W. Bickley,$^{1}$\thanks{E-mail: rbickley@uvic.ca}
Sara L. Ellison,$^{1}$
Mara Salvato,$^{2}$
\newauthor
Samir Salim,$^{3}$
David R. Patton,$^{4}$
Andrea Merloni,$^{2}$
\newauthor
Shoshannah Byrne-Mamahit,$^{1}$
Leonardo Ferreira,$^{1}$
Scott Wilkinson$^{1}$
\\
$^{1}$Department of Physics and Astronomy, University of Victoria, Victoria, British Columbia V8P 1A1, Canada\\
$^{2}$Max-Planck-Institut für extraterrestrische Physik (MPE), Gießenbachstraße 1, D-85748 Garching bei München, Germany\\
$^{3}$Department of Astronomy, Indiana University, Bloomington, Indiana 47405, USA\\
$^{4}$Department of Physics and Astronomy, Trent University, 1600 West Bank Drive, Peterborough, ON K9L 0G2, Canada\\
}

\date{Accepted XXX. Received YYY; in original form ZZZ}

\pubyear{2024}

\begin{document}
\label{firstpage}
\pagerange{\pageref{firstpage}--\pageref{lastpage}}
\maketitle

\begin{abstract}
Major mergers between galaxies are predicted to fuel their central supermassive black holes (SMBHs), particularly after coalescence. However, determining the prevalence of active galactic nuclei (AGNs) in mergers remains a challenge, because AGN diagnostics are sensitive to details of the central structure (e.g., nuclear gas clouds, geometry and orientation of a dusty torus) that are partly decoupled from SMBH accretion. X-rays, expected to be ubiquitous among accreting systems, are detectable through non-Compton-thick screens of obscuring material, and thus offer the potential for a more complete assessment of AGNs in mergers. But extant statistical X-ray studies of AGNs in mergers have been limited by either sparse, heterogeneous, or shallow on-sky coverage. We use new X-ray observations from the first SRG/eROSITA all-sky data release to characterize the incidence, luminosity, and observability of AGNs in mergers. Combining machine learning and visual classification, we identify 923 post-mergers in Dark Energy Camera Legacy Survey (DECaLS) imaging and select 4,565 interacting galaxy pairs (with separations <120 kpc and mass ratios within 1:10) from the Sloan Digital Sky Survey. We find that galaxies with X-ray AGNs are 2.0$\pm$0.24 times as likely to be identified as post-mergers compared to non-AGN controls, and that post-mergers are 1.8$\pm$0.1 times as likely to host an X-ray AGN as non-interacting controls. A multi-wavelength census of X-ray, optical, and mid-IR-selected AGNs suggests a picture wherein the underlying AGN fraction increases during pair-phase interactions, that galaxy pairs within \textasciitilde20 kpc become heavily obscured, and that the obscuration often clears post-coalescence.
\end{abstract}




\section{Introduction}
\label{Introduction}

Major mergers between galaxies of similar mass punctuate the otherwise gradual evolution of isolated galaxies with short epochs of intense change. The tidal influence of merging companions disrupts galaxy morphologies, resulting in visually distinct features: stellar streams connecting interacting galaxies, asymmetric tidal tails, and stellar shells produced by coherent torques exerted near the end of the merger (\citealp{2010MNRAS.401.1043D}; \citealp{2015ApJS..221...11K}; \citealp{2016MNRAS.461.2589P}; \citealp{2017MNRAS.464.4420S}). With each close pericentric passage, the participant galaxies exert strong tidal torques on one another, disrupting the angular momentum of gas and channeling it towards the galaxies' centres. Merging galaxy pairs in both simulations (\citealp{1972ApJ...178..623T}; \citealp{2005MNRAS.361..776S}; \citealp{2008AN....329..952D}; \citealp{2019MNRAS.490.2139R}; \citealp{2019MNRAS.485.1320M}; \citealp{2020MNRAS.494.4969P}) and observations (\citealp{2004MNRAS.355..874N}; \citealp{2008AJ....135.1877E, 2011MNRAS.418.2043E, 2013MNRAS.435.3627E,2019MNRAS.487.2491E}; \citealp{2012MNRAS.426..549S}; \citealp{2013MNRAS.433L..59P}; \citealp{Lackner_2014};  \citealp{2014MNRAS.441.1297S}; \citealp{2014MNRAS.437.2137S}; \citealp{2015HiA....16..326K}; \citealp{10.1093/mnras/stw2620}; \citealp{2018PASJ...70S..37G}) show elevated star formation rates (SFRs), star-forming fractions, and SMBH accretion rates as a result of the elevated central gas concentrations brought on by the merger. Simulation-based (\citealp{2020MNRAS.493.3716H}; \citealp{2023MNRAS.519.4966B}) and observational studies (\citealp{10.1111/j.1365-2966.2011.20179.x}; \citealp{2014MNRAS.441.1297S}; \citealp{2019MNRAS.482L..55T}; \citealp{2022MNRAS.514.3294B,2023MNRAS.519.6149B}; \citealp{2023ApJ...944..168L}) have found that when merging galaxies coalesce into a sovereign post-merger system (and for a short time thereafter), the influence of the merger on star formation and SMBH activity is strongest. After some time, through either exhaustion of the gas supply by star formation, heating by the AGN, or removal of the gas via ejective feedback, mergers have also been predicted (e.g., \citealp{2021MNRAS.504.1888Q,2023MNRAS.519.2119Q}; \citealp{2022MNRAS.515.1430D}) and observed (\citealp{2013MNRAS.435.3627E}; \citealp{2022MNRAS.517L..92E}; \citealp{2023MNRAS.523..720L}) to truncate star formation, and simulations suggest that post-merger AGN fractions (e.g., \citealp{2023MNRAS.519.4966B}) once again become consistent with non-interacting galaxies after \textasciitilde1.5 Gyr. Post-mergers therefore represent the integrated effect of a merger event on a galaxy's evolutionary trajectory.

The connection between galaxy mergers and AGNs has been demonstrated observationally using multiple AGN selection techniques including the use of optical emission lines and mid-IR colours (\citealp{1985AJ.....90..708K}; 
\citealp{2007MNRAS.375.1017A}; \citealp{2007AJ....134..527W}; \citealp{2014MNRAS.441.1297S}; \citealp{10.1093/mnras/stw2620}; \citealp{2018PASJ...70S..37G}; \citealp{2020A&A...637A..94G}; \citealp{2020MNRAS.499.2380S}). These various AGN selection methods follow naturally from the unified AGN model (see the review by \citealp{2018ARA&A..56..625H}), assuming that rapidly accreting AGNs universally host a UV-bright accretion disk, a parsec-scale broad-line region (BLR) orbiting at thousands of km/s, and a hot corona where UV photons from the accretion disk are excited to X-ray energies via inverse Compton scattering interactions with relativistic electrons \citep{1993ARA&A..31..717M,1994ApJ...432L..95H,2016MNRAS.458..200W,2019MNRAS.487..667C}. Direct detection of the BLR (as in a type 1 AGN), the time-variable UV / optical blue blackbody radiation from the accretion disk, or X-rays from the corona in excess of the expected luminosity from X-ray binaries in young stellar populations (\citealp{2003A&A...399...39R}; \citealp{2019ApJS..243....3L}) are unambiguous evidence for the presence of an actively accreting SMBH in a galaxy.

Different AGN selection methods are prone to varying degrees of incompleteness (e.g., \citealp{2016MNRAS.457..110M}; \citealp{2019ApJ...876...12A}) and studies investigating the merger-AGN connection naturally inherit the biases associated with the AGN criteria used. Narrow-line optical (e.g., \citealp{2011ApJ...729..141B}), mid-IR (e.g., \citealp{2013ApJ...772...26A}) and X-ray selections (e.g., a sample from XMM-XLL North in \citealp{2016MNRAS.457..110M}) only overlap partially. Even with deep X-ray observations (e.g., studied in a sample complete to $\mathrm{L_{X}}$>$10^{42}$ in \citealp{2019ApJ...876...12A}), obscuring material leads to narrow-line AGNs that are undetectable in the X-ray (which are either intrinsically faint or heavily obscured). In the opposite case, emission from faint narrow-line AGNs can be dominated by nebular ionization from star-forming regions, giving rise to a population of X-ray bright AGNs with the optical appearance of purely star-forming galaxies.

Further complicating the issue, mergers impede their own study. Particularly in late-stage galaxy pairs and post-mergers, the same inflows that fuel AGNs also increase the amount and covering fraction of obscuring nuclear dusty gas (\citealp{2017Natur.549..488R}; \citealp{2018MNRAS.478.3056B}). At the time of coalescence, competing effects influence the final observability of AGNs: if the dust column dominates over the SMBH luminosity, post-mergers will be observed as heavily obscured or Compton-thick. If the SMBH is accreting rapidly enough to dominate over the density of obscuring material, however, the dusty gas can be cleared away (increasing the outflow rate from the nucleus by a factor of \textasciitilde10 as in  \citealp{2016MNRAS.458..816H}), and the observational evidence for SMBH accretion becomes plain (e.g., as modelled in \citealp{2008MNRAS.385L..43F} and tested in \citealp{2022ApJ...938...67R}). Classically, circum-nuclear blowout is associated with late stage galaxy mergers (\citealp{1996ARA&A..34..749S}).

X-ray AGN selection is complementary to optical and mid-IR selections, and is potentially less affected by the geometry of obscuring material. Except in heavily obscured cases, X-ray luminosity characterizes the state of the SMBH reliably, since the BLR is often obscured by dust, and the observability of the narrow-line region and the torus is conditional. Previous X-ray observations have been impeded by either sparse or heterogeneous on-sky coverage, or shallow depth. X-ray catalogues have therefore limited the prospects for a statistical analysis of X-ray AGNs in galaxy mergers, although some constraints have been placed after careful treatment of detections with heterogeneous signal to noise ratios (S/N). Statistical studies of galaxy pairs (e.g., \citealp{2020ApJ...900...79H,2023ApJ...943...50H}; \citealp{2021MNRAS.504..393G}; \citealp{2021MNRAS.506.5935R}; \citealp{2023ApJ...949...49H}) have indicated that pair-phase interactions can trigger X-ray AGNs. Most studies have investigated the incidence rate of X-ray AGNs in galaxy pairs identified in the optical (e.g., \citealp{2020ApJ...900...79H,2023ApJ...943...50H}; \citealp{2021MNRAS.504..393G}) or the infrared (e.g., \citealp{2021MNRAS.506.5935R}). The relative frequency of X-ray AGNs in pairs and isolated galaxies is therefore still uncertain. X-ray AGNs in close galaxy pairs (i.e., with projected separations below 20 kpc) seem to be relatively rare (as suggested by \citealp{2023ApJ...949...49H}). There is some tension regarding the cause of this rarity, with some (e.g., \citealp{2016ApJ...825...85K}; \citealp{2021MNRAS.506.5935R}; \citealp{2021MNRAS.504..393G}) suggesting that gas inflows in the final stages of mergers give rise to heavy or Compton-thick AGN obscuration, while others (e.g., \citealp{2023ApJ...943...50H}) suggest that AGN feedback in late stage galaxy pairs can prevent further accretion onto the SMBH. Strengthening the observed connection between late stage galaxy mergers, SMBH accretion, and obscuration, luminous obscured SMBHs have been shown to host an excess of late-stage mergers (e.g., a significant excess of 16 in \citealp{2018Natur.563..214K}). Statistical studies of post-mergers indicate that X-ray AGNs are triggered by coalescence (e.g., \citealp{2011ApJ...739...57K}), and the excess of X-ray AGNs in post-mergers compared to non-interacting control galaxies appears to be a factor of \textasciitilde2 (\citealp{2020MNRAS.499.2380S}; \citealp{2023ApJ...944..168L}). However, the particular value and statistical significance of the excess is highly sensitive to the merger selection, AGN criteria, and sample size studied. The observational ubiquity of X-ray AGNs in galaxy mergers, as well as that of mergers amongst X-ray AGNs, is uncertain on account of diverse merger samples, small X-ray catalogues, and heterogeneous AGN criteria. The relative roles of intrinsic luminosity and obscuration in producing AGNs with a given $\mathrm{L_{X}}$ as a function of merger stage are even more opaque.

The extended ROentgen Survey with an Imaging Telescope Array (eROSITA, the X-ray survey used in this work) all-sky survey (\citealp{2021A&A...647A...1P}; \citealp{2024A&A.XXX.XXXXM}) offers an opportunity for substantial improvement on the completeness and homogeneity of other X-ray catalogues. The first all-sky data release from eROSITA, eRASS1, surveys the entire sky at approximately uniform depth. eRASS1's ability to detect AGNs is sensitive to obscuration, and we anticipate that Compton-thick AGNs (typical column densities $\mathrm{N_{H}}$>$10^{24}$/cm$^{2}$ e.g., in \citealp{2019A&A...629A.133C}) and some Compton-thin obscured AGNs (where the obscuration is still sufficient to suppress soft X-rays, $\mathrm{N_{H}}\gtrsim10^{22}$/cm$^{2}$) will not be detected (\citealp{2022A&A...661A...5L}). eROSITA is therefore most sensitive to unobscured AGNs (\citealp{2024arXiv240117306W}), and the relative contributions of intrinsic luminosity and obscuration are not observationally discernible in the eRASS1 sample.

Excepting obscured cases, we can use eRASS1 to homogeneously detect X-ray AGNs in the low-$z$ Universe. We note that the eRASS1 X-ray AGN selection targets only the top of the dynamic range of accretion rates, and detected luminosities can be attenuated in galaxies with dense obscuring columns. Multiple diagnostics are often considered together to study a broader range of AGNs (e.g., \citealp{2023ApJ...944..168L}; \citealp{2023MNRAS.519.6149B}), and we adopt this approach by supplementing our X-ray AGN sample with a multi-wavelength analysis (Section~\ref{Results}).

In this work, we will study the influence of merger events on the incidence, observed luminosity, and observability of AGNs. Thanks to the statistically large samples of X-ray detections (via eRASS1) and post-mergers (using a hybrid machine vision plus visual classification approach), plus multi-wavelength legacy data from SDSS DR7 (for optical characterization) and unWISE\footnote{unwise.me} photometry (for mid-IR; \citealp{2019ApJS..240...30S}), we are able to outline observationally each of the physical effects discussed above: triggering of X-ray AGNs in galaxy pairs and post-mergers, SMBH luminosities in mergers, and the likelihood and type of AGN obscuration as a function of merger stage.

In Section~\ref{Methods}, we describe our data products: galaxy post-merger, pair, and isolated (control) samples, optical, mid-infrared, and X-ray AGN selection criteria, and control matching methodology. We next present our results, characterizing the incidence, luminosities, and multi-wavelength observability of X-ray AGNs in interacting galaxies (Section~\ref{Results}). We discuss the implications of our results for the completeness and accuracy of future AGN studies (Section~\ref{Discussion}), and finally summarize the results of this study (Section~\ref{Summary}). We assume cosmological parameters ($\Omega_{\mathrm{m0}}$ = 0.3, $\Omega_{\Lambda0}$ = 0.7, $h$= 0.7) when calculating luminosity distances, projected separations between interacting galaxies, and any other cosmology-dependent quantities appearing in this work.

\section{Methods}
\label{Methods}

The work presented here is mainly an X-ray-focused study of pre-mergers and completed galaxy mergers, supplemented by additional multi-wavelength observations. In this Section, we describe the parent sample of galaxies, as well as the selection methods for post-mergers, galaxy pairs, control pool galaxies, and multi-wavelength AGNs.

\subsection{Galaxy samples}
\label{Galaxy samples}

\subsubsection{Parent catalogue}
\label{Parent catalogue}
The results of this work depend on homogeneous availability of our merger, X-ray, and multi-wavelength AGN data products for the entire galaxy sample studied. The parent catalogue is outlined as meeting the following criteria:

\begin{itemize}
   \item Available in SDSS DR7 with a redshift, and assigned a spectroscopic class equal to 2 (galaxy), 3 (quasar), or 4 (high-$z$ quasar) by the SDSS spectral classification pipeline.
   \medskip
   \item $z$>0.005 to avoid any Galactic objects misclassified by the SDSS pipeline, and $z$<0.3 to limit the study to the $z$ domain where our merger classifications are homogeneously reliable (see \citealp{2021MNRAS.504..372B})
   \medskip
   \item Within the eROSITA (German consortium\footnote{The German consortium has access to half of the all-sky survey, with the Galactic longitudes given in the text marking the boundary.}) footprint with Galactic longitudes between $180^{\circ}$<$l$<$360^{\circ}$.
   \medskip
   \item Dark Energy Camera Legacy Survey (DECaLS, \citealp{2019AJ....157..168D}) $r$-band imaging available, taken with the Dark Energy Camera for the Legacy Survey\footnote{www.legacysurvey.org/dr10/description/} data release 10 (LS10). LS10 covers the entire eROSITA-DE footprint, provides deep $i$ band, and deeper WISE imaging compared to the 9th data release (\citealp{2019AJ....157..168D}). DECaLS $r$-band imaging is used in this work to search for the morphological signatures of a recent merger event.
\end{itemize}

There are 265,371 galaxies meeting these criteria. The results presented in this work require merger and control samples that are well matched on redshift (which is available by definition for the entire catalogue via SDSS DR7), and stellar mass. A uniform stellar mass estimate is not available for the entire catalogue, since the MPA-JHU\footnote{wwwmpa.mpa-garching.mpg.de/SDSS/DR7/} (\citealp{2004MNRAS.351.1151B}) stellar mass catalogue is constructed mainly for galaxies with narrow emission lines. Since many X-ray AGNs are observed to have broad lines, using the MPA-JHU mass catalogue would lead to an incomplete census of AGNs in mergers. Instead, we generate new stellar mass estimates for use throughout this study.

Stellar masses are computed following the method of \citet{2016ApJS..227....2S}, but using only optical model magnitudes in the $griz$ bands. \citet{2016ApJS..227....2S} allowed for additional photometry (SDSS $u$, UV from GALEX, \citealp{2005ApJ...619L...1M}, and mid-IR from WISE) to be considered when performing SED fitting for stellar mass and SFR estimates. However, we exclude SDSS $u$ and UV photometry on account of potential contamination by emission from the AGN accretion disk, and also exclude mid-IR photometry from the stellar mass fits in order to avoid biases from AGN-heated dust. We follow \citet{2016ApJS..227....2S} and use the SED fitting code CIGALE (\citealp{2009A&A...507.1793N}; \citealp{2016A&A...591A...6B}) to fit grids of synthetic stellar continuum templates from \citet{2003MNRAS.344.1000B} and star formation emission line templates to produce probability density functions (PDFs) for stellar mass (M$_\mathrm{\star}$). A posteriori, we inspect the distribution of M$_\mathrm{\star, new}-$M$_\mathrm{\star, MPA-JHU}$ and find that \textasciitilde88\% of galaxies listed in both MPA-JHU and the new M$_\mathrm{\star}$ catalogue agree within 0.1 dex. The median difference is \textasciitilde0.03 dex (with the new mass estimate being slightly higher on average), and the standard deviation of the mass difference distribution is \textasciitilde0.08 dex. The broad agreement between the new stellar mass estimates and the MPA-JHU catalogue suggests that the new M$_\mathrm{\star}$ catalogue offers an important improvement in the availability of stellar mass estimates without sacrificing precision. For the interested reader, we characterize the masses determined if the $u$ and UV bands are included in Appendix~\ref{A1}.

Photometry flags are raised when the SED fitting cannot account for an unusual colour for an object in the catalogue, typically due to imaging artifacts in SDSS. Dropping the sources with photometry flags only reduces the sample by 0.05\%, and so has no impact on our results. To be included in the parent catalogue, we require that galaxies have $\mathrm{log(M_{\star}/M_{\odot})}$>8 and no photometry flags raised during stellar mass fitting. The mass cut is chosen to be inclusive of the low-mass wings of the eRASS1-detected sample. There are 264,820 galaxies in the parent catalogue with spectroscopic redshifts in the range $0.005<z<0.3$ and $griz$-determined stellar masses in the range $8<\mathrm{log(M_{\star}/M_{\odot})}<12.75$.

\subsubsection{Post-merger sample}
\label{Post-merger sample}

Having defined our parent sample of galaxies, we can now proceed with the identification of the merger sample. Post-mergers in this work are identified using the hybrid classification scheme introduced in \citet{2021MNRAS.504..372B}, wherein a convolutional neural network (CNN) was trained on post-mergers and isolated controls from the TNG100-1 run of the IllustrisTNG simulation (\citealp{2018MNRAS.475..648P}; \citealp{2018MNRAS.475..676S}; \citealp{2018MNRAS.480.5113M}; \citealp{2018MNRAS.475..624N}; \citealp{2018MNRAS.477.1206N}) with Canada-France Imaging Survey (CFIS, part of the UNIONS\footnote{www.skysurvey.cc} consortium) $r$-band sky background and survey-accurate realism added. Mock galaxy observations were generated by adapting the method of \citet{2019MNRAS.490.5390B} for CFIS. The CNN was next validated on like-generated simulated galaxy data before being used to classify real CFIS galaxies in \citet{2022MNRAS.514.3294B} and \citet{2023MNRAS.519.6149B}. The CNN makes a decimal prediction between zero and one representing the likelihood that the galaxy is a post-merger, hereafter referred to as $p(x)$. Visual inspection is then used to confirm merger status for galaxies with high CNN-predicted probabilities.

For this work, we use the exact CNN model and weights as in \citet{2022MNRAS.514.3294B,2023MNRAS.519.6149B} to classify galaxies in DECaLS $r$-band images. Although the model was originally trained and used to classify CFIS imaging, DECaLS has substantially better overlap with eROSITA (CFIS has a minimum declination coverage of 30$^{\circ}$). DECaLS is not as deep (with a 5$\sigma$ $r$-band point-source depth of 23.9 mag compared to CFIS's 25.0 mag) and has coarser pixel-scale resolution by a factor of \textasciitilde2. The DECaLS pixel scale determines the effective signal to noise ratio in the images, resulting in higher effective surface brightness sensitivity in DECaLS compared to CFIS (by about 40\%). It is necessary to test whether the original CNN is appropriate to use on DECaLS imaging. We therefore prepared 100-kpc-square cutouts of the subset of 16,025 galaxies (69 of which are post-mergers in the \citealp{2022MNRAS.514.3294B} and \citealp{2023MNRAS.519.6149B} catalogues) imaged by both DECaLS and CFIS in the $r$-band in the thin stripe of declination between 30$^{\circ}$ and 32.375$^{\circ}$, normalized both sets of images using the method in \citet{2021MNRAS.504..372B}, and compared the CNN's classifications between the two surveys. Using a CNN $p(x)$ criterion of 0.6 to select post-mergers in DECaLS imaging, we find that we recover \textasciitilde 68\% (47 galaxies) of the visually confirmed post-mergers from \citet{2023MNRAS.519.6149B} in the survey overlap. The other 32\% (22 galaxies) are lost (below $p(x)=0.6$). However, we find that a significant number of new visually convincing post-mergers (some 30 galaxies in the overlap) are recovered in DECaLS imaging that were not found by the CNN in the CFIS imaging. Mergers "discovered" in DECaLS and not in CFIS by the same CNN are sometimes due to CFIS imaging artifacts (which are often absent in DECaLS), but in other cases the reason for a given merger's new classification in DECaLS is not visually obvious. In some uncertain cases, the "discovery" of new post-mergers in DECaLS could be owed to the higher effective spatial S/N of DECaLS imaging, since the CNN's performance is sensitive to the brightness of low-surface brightness regions of galaxy images (see Figure A1 of \citealp{2022MNRAS.514.3294B}). The above experiment demonstrates two relevant effects. First, merger recovery is somewhat stochastic and not simply a function of image depth (this is explored in detail in Ferreira et al. in prep, in which multiple networks are tested in a jury system). Second, we expect that a statistically compelling sample of mergers can be identified from the DECaLS imaging using the CNN we already have in hand. We emphasize that completeness is relatively unimportant for our study, but purity is. This is because our goal is to characterize the AGN population in a bona-fide sample of mergers, rather than have a complete sample of mergers themselves. While the choice of $p(x)>0.6$ to identify the sample for visual inspection is arbitrary, our aim is to identify a large sample of visually-confirmed post-mergers in DECaLS, and the statistics of the cross-survey validation experiment described above indicate that $p(x)>0.6$ will allow us to do so.

Having demonstrated that our existing CNN can be used to recover a post-merger sample in DECaLS imaging, we next produce 100-kpc-square DECaLS-$r$ cutouts of the \textasciitilde265,000 galaxies in the parent sample, resize them to 138$\times$138 pixels\footnote{We use 138$\times$138 pixels because this was the size of a CFIS $r$-band image with a diameter of 100kpc at the median redshift of the \citet{2022MNRAS.514.3294B} sample. Consequently, the \citet{2022MNRAS.514.3294B} CNN takes in images of this size.}, normalize them in linear fashion between $0-1$, and classify them with the CNN. As with CFIS, some images contain artifacts (e.g., saturated foreground stars, missing pixels) that partially or completely mask the morphology of the galaxy. We classify these objects anyway, since the CNN was trained on images with artifacts, and we will have an opportunity to remove any objects with prominent artifacts masquerading as post-mergers in the visual classification stage of our merger effort (in practice, such objects are rare).

Finally, we conduct a visual classification effort in order to remove false positives from the sample. 4,093 galaxies received CNN post-merger predictions $p(x)$>0.6, and they were inspected by the author RWB. A consultation for select ambiguous cases was completed (with authors DRP and SLE), and the conclusions of the committee were taken into account during a final check of the sample by RWB. For post-mergers only, galaxies were removed from the final sample if they:

\begin{itemize}
	\item{lacked visually convincing post-merger morphology, or}
	\medskip
	\item{were obscured by one or more imaging artifacts such that the morphology could not be discerned, or}
	\medskip
	\item{had a nearby companion in DECaLS imaging that might have been interacting, or}
	\medskip
	\item{had two or more nuclei, suggesting pre-merger rather than post-merger status.}
\end{itemize}

\begin{figure*}
\includegraphics[width=\textwidth]{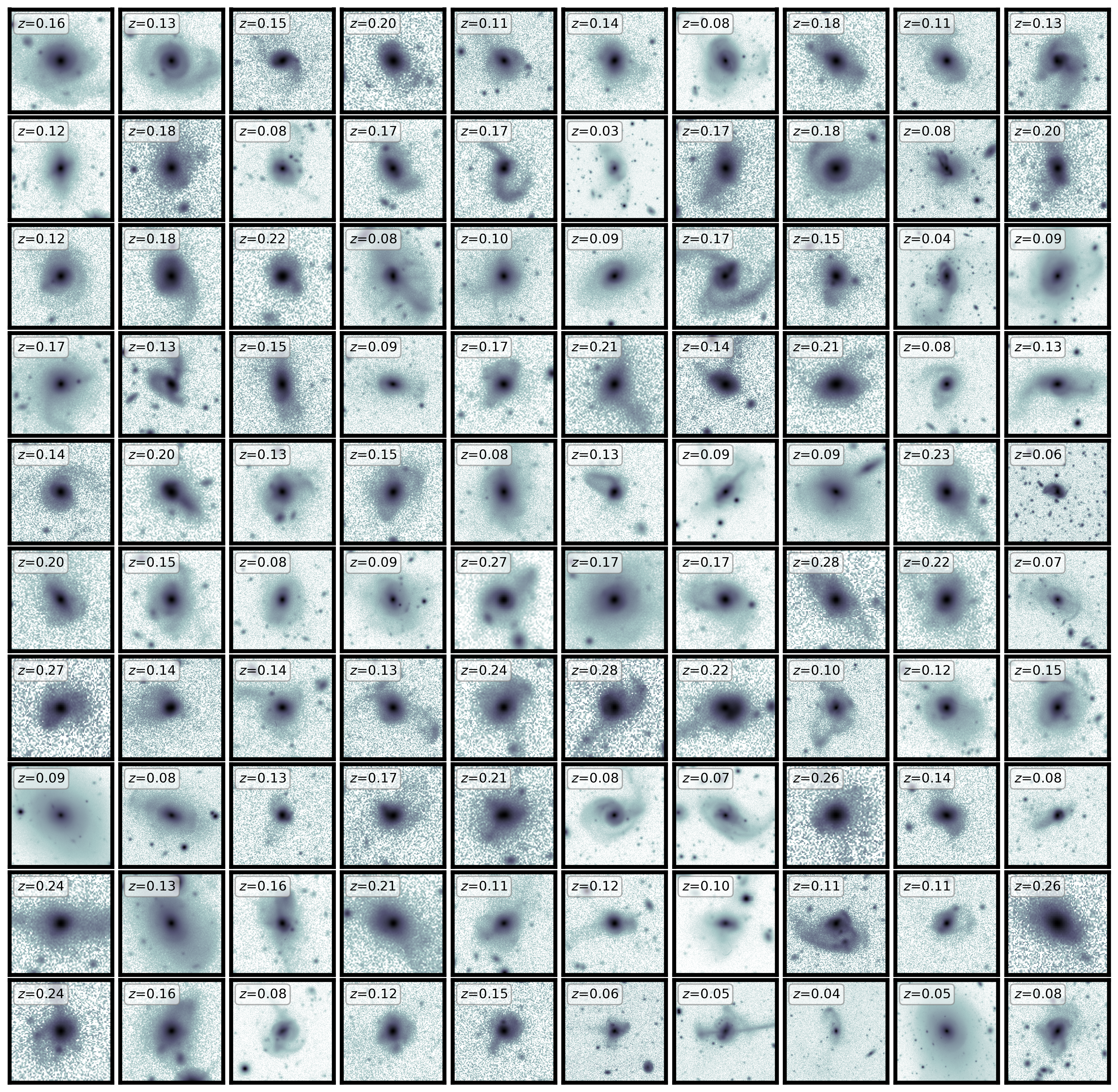}
\caption{Mosaic of 100 $r$-band DECaLS images of post-merger galaxies selected at random from the visually confirmed sample of 923 identified for this work, annotated with the SDSS DR7 spectroscopic $z$. Cutouts are 100 kpc on a side. Images are shown with the mean subtracted, normalized by the standard deviation, and converted to log scale in order to highlight the low-surface-brightness features that earned them post-merger classifications.}
\label{fig:mosaic}
\end{figure*}

Of the 4,093 galaxies inspected, 923 were confirmed as having the unambiguous characteristics of recent coalescence; 100 random examples are shown in Figure~\ref{fig:mosaic}. We note that the visual confirmation fraction (23\%) from the DECaLS post-merger search is lower than that of \citet{2022MNRAS.514.3294B} (36\%). We attribute the lower confirmation rate to the fact that the signatures of a recent merger are visually weaker (and therefore, less convincing) in DECaLS compared to CFIS due to the \textasciitilde1 mag difference in sensitivity. As a result, a greater fraction of predicted post-mergers had somewhat ambiguous merger features in DECaLS, and were rejected during the visual classification effort. The sensitivity difference between the two surveys likely means that the post-merger remnants selected for this work have somewhat more dramatic morphologies (following a merger with a higher mass ratio or more disruptive orbital parameters, or with less time elapsed since coalescence). Nonetheless, at the end of this process we have a large, pure sample of post-mergers in which AGN incidence can be studied.

\subsubsection{Pair sample}
\label{Pair sample}

We identify a bespoke sample of spectroscopic galaxy pairs in the parent sample (with shared overlap between SDSS DR7, DECaLS-$r$, eRASS1) using the on-sky coordinates of the galaxies in our parent catalogue, their spectroscopic redshifts from SDSS, and the cosmological parameters assumed in this work. We follow generally the method of \citet{2016MNRAS.461.2589P} with small modifications in order to identify a more complete catalogue meeting specific criteria relevant to this work.

For each galaxy in the parent catalogue, we find the closest companion (with the smallest projected separation, hereafter $\mathrm{r_{p}}$) whose stellar mass is within a factor of 10 and whose line-of-sight velocity difference ($\mathrm{\Delta}$v) is less than 300 km/s. Our mass ratio criterion is motivated by the fact that simulated post-mergers with 0.1<$\mu$<10 were selected to train the CNN used for post-merger classifications in this work. The pairs thus selected plausibly represent pre-coalescence analogues to the post-merger sample. Any galaxy with such an interacting companion closer than 120 kpc in $\mathrm{r_{p}}$ is labelled as a member of a pair for this work. A projected separation of 120 kpc is (at least) twice the separation where previous studies (e.g., \citealp{2013MNRAS.435.3627E}; \citealp{2023MNRAS.519.6149B}) have found an AGN excess and therefore represents a sufficient buffer. Finally, we require that our pairs not be included in the post-merger sample described in Section~\ref{Post-merger sample}. By these criteria, there are 15,485 galaxies with a companion with $\mathrm{r_{p}}$<120 kpc.

In addition to the sample of closest companions described above, we separately define a sample of "mutual galaxy pairs" (MGPs). MGPs are sets of exactly two galaxies defined by mutual companionship (i.e., galaxy $A$'s tabulated companion in the catalogue is galaxy $B$, and galaxy $B$'s companion is $A$). Mutual pairs are distinct from galaxies belonging to more complex spatial families on the sky (e.g., where galaxy $B$'s nearest neighbour is a third galaxy, $C$, instead of $A$). The sample of MGPs is required in our analysis of AGNs in eRASS1, since the resolution of that survey is insufficient to resolve individual galaxies in a close pair. Our X-ray analysis is therefore done on a pairwise basis (see Section~\ref{Pairwise treatment of X-ray detections in galaxy pairs}). From the sample of close companions described above, we identify 4,565 MGPs (i.e., 9,130 galaxies) with projected separations <120 kpc.

\subsubsection{Control pool}
\label{Control pool}

Post-merger and pair galaxies are compared to samples of matched controls throughout this work. Control galaxies are selected from those in the parent sample that have no close companion within 1000 km/s in $\Delta$v, 0.1<$\mu$<10 in mass ratio, and $\mathrm{r_{p}}$<120 kpc. We use these criteria because galaxies with $\Delta$v>1000 km/s are likely not destined to merge, or at different $z$ entirely. We select control galaxies that are separated from their companions by at least 120 kpc in order to compare our merger samples to controls that are evolving in effective isolation compared to our sample of galaxy pairs. We use the same mass ratio criteria as for galaxy pairs in order to avoid a mass ratio asymmetry in our analysis. Control galaxies must also have CNN $p(x)<0.1$ predictions from DECaLS imaging.

In order to construct a control sample for each galaxy in the pair and post-merger samples, we begin by identifying the best simultaneous match in $z$ and $\mathrm{M_{\star}}$ in the control pool. We create normalized log($\mathrm{M_{\star}}$) and $z$ statistics for the control pool by subtracting their mean value and dividing the result by their standard deviation. We then use the normalized log($\mathrm{M_{\star}}$) and $z$ to create a KDTree\footnote{docs.scipy.org/doc/scipy/reference/generated/scipy.spatial.KDTree.html}, an algorithm which can be used to efficiently identify a galaxy's nearest neighbours in parameter space. We apply the same normalization to the merger sample, and identify the best control galaxy (without replacement) for each merger. We repeat the process, allowing for as many as 10 unique controls to be matched to each merger. After each batch of controls is matched, we apply a two-sample Kolmogorov-Smirnov (K-S, \citealp{smirnov1948}) test to the $\mathrm{M_{\star}}$ and $z$ statistics of the merger and control samples. The K-S test is designed to estimate the likelihood that two samples were drawn from the same parent distribution. In this context, high K-S $p$-values indicate that the control matching is working well to select control samples that are statistically indistinct from the mergers in $\mathrm{M_{\star}}$ and $z$. If the K-S $p$-value falls below 0.9 for either galaxy statistic, we reject the most recent batch of controls and terminate control matching. Typically, 10 batches of controls are successfully matched to the merger (pair, MGP, and post-merger) samples. The most poorly matched control galaxy in this study is offset in log($\mathrm{M_{\star}}$) by 0.7 dex, but more than 99\% of galaxies are matched to controls within 0.02 dex in log($\mathrm{M_{\star}}$). The most poorly matched control galaxy in $z$ is offset by 0.01, and more than 99\% of galaxies are matched to controls within 0.002 in $z$. We find that matching controls on environmental parameters (environment density, neighbour distances) in addition to $\mathrm{M_{\star}}$ and $z$ does not qualitatively change our results, even though the methods of selection for post-mergers and galaxy pairs means that they reside in somewhat sparse (for the post-mergers) and dense (for the pairs) environments relative to the average control galaxy. Requiring a control pool that is even more isolated (e.g., with $\mathrm{r_{p}}>200$ kpc) also has little influence on the conclusions of this work.

\subsection{AGN samples}
\label{AGN samples}

\subsubsection{X-ray sample}
\label{X-ray sample}

The X-ray fluxes used in this work are taken from the first data release of the eROSITA all-sky survey (\citealp{2021A&A...647A...1P}; \citealp{2024A&A.XXX.XXXXM}). eRASS1 surveys the entire sky down to a flux limit of approximately $5\times10^{-14}$erg/s/cm$^{-2}$ in the 0.2$-$5 keV band, corresponding to limiting observed luminosities of $\mathrm{L_{X}}=10^{42}$, $10^{42.7}$, and $10^{43.1}$ erg/s at $z=0.1$, 0.2, and 0.3, respectively. The typical eROSITA point spread function (PSF) half-energy width\footnote{The half-energy width is a cumulative version of the Full Width at Half Max (FWHM).} (HEW) is \textasciitilde30 arcsec. X-ray detections considered in this work have detection likelihoods greater than 5, and spurious detections and extended sources have been removed in advance. Since highly luminous AGNs are rare, a relatively large proportion of the sample is likely intrinsically faint, and thereby close to the eRASS1 detection limit (i.e., observed in eRASS1 with only a few counts). Consequently, the most common effect of flux suppression is to remove sources from the sample. We therefore expect that heavily obscured AGN are more often absent from the X-ray sample than they are detected through the obscuration. We use only observed fluxes and luminosities in this work since we lack the information to measure obscuration and extract intrinsic luminosities.

Optical counterparts for the X-ray sources are identified probabilistically from multi-band Legacy Survey Data Release 10 (LS10) imaging, based on the Bayesian method outlined in \citet{2018MNRAS.473.4937S}. The counterparts used in this work are presented in Salvato et al. (in prep). Candidate counterparts within the positional uncertainty of the X-ray source are each assigned a probability $p_{i}$ of being the correct counterpart of the X-ray detection, and are included in this work if the source with the maximum $p_{i}$ for a given X-ray detection appears in the parent catalogue described in Section~\ref{Parent catalogue}. Maximum likelihood fluxes from the eRASS1 catalogue are converted to luminosities based on the SDSS spectroscopic redshifts of the parent sample galaxies. eRASS1 is able to detect AGNs with observed luminosities >$10^{43.1}$ erg/s for any galaxy in this study (with $z$<0.3). For a typical redshift ($z$=0.1) in our sample, eRASS1 is more than an order of magnitude more sensitive. Interacting galaxy pairs, particularly at small projected separations, are quite likely to be crowded (with the interacting galaxies not individually resolved) within the eROSITA PSF, and have their X-ray emission blended (in the case where more than one source is contributing X-rays to the detection). We present a new analytic method to account for this effect in Section~\ref{Pairwise treatment of X-ray detections in galaxy pairs}.

AGNs are not the only source of X-ray emission in galaxies, with some binary star systems also producing significant emission (\citealp{2003A&A...399...39R}; \citealp{2019ApJS..243....3L}). However, at the sensitivity of eRASS1, the X-ray sources detected and associated with galaxies in the parent catalogue are highly luminous, and therefore unlikely to be the product of X-ray binaries in young stellar populations in the host galaxies. For example, only six of the 3,718 (0.2\%) eRASS1-detected galaxies have MPA-JHU SFRs in excess of the \citet{2003A&A...399...39R} $\mathrm{L_{X}}$ criterion adapted by \citet{2019ApJ...876...12A}. Only one of the six cases is a post-merger, and it is an AGN by both our mid-IR (see Section~\ref{Mid-IR AGN sample}) and narrow-line (Section~\ref{NLAGN sample}) AGN criteria. We therefore count any X-ray source in our sample as an X-ray AGN, and note that removing the small number of galaxies with anomalously high SFRs from the X-ray AGN sample does not impact our results. There are 3,718 galaxies with X-ray detections in our parent catalogue. Of the X-ray AGN detected galaxies, 58 are post-mergers, 232 are in an interacting pair, and 49 are in a MGP.

\begin{figure}
\includegraphics[width=\columnwidth]{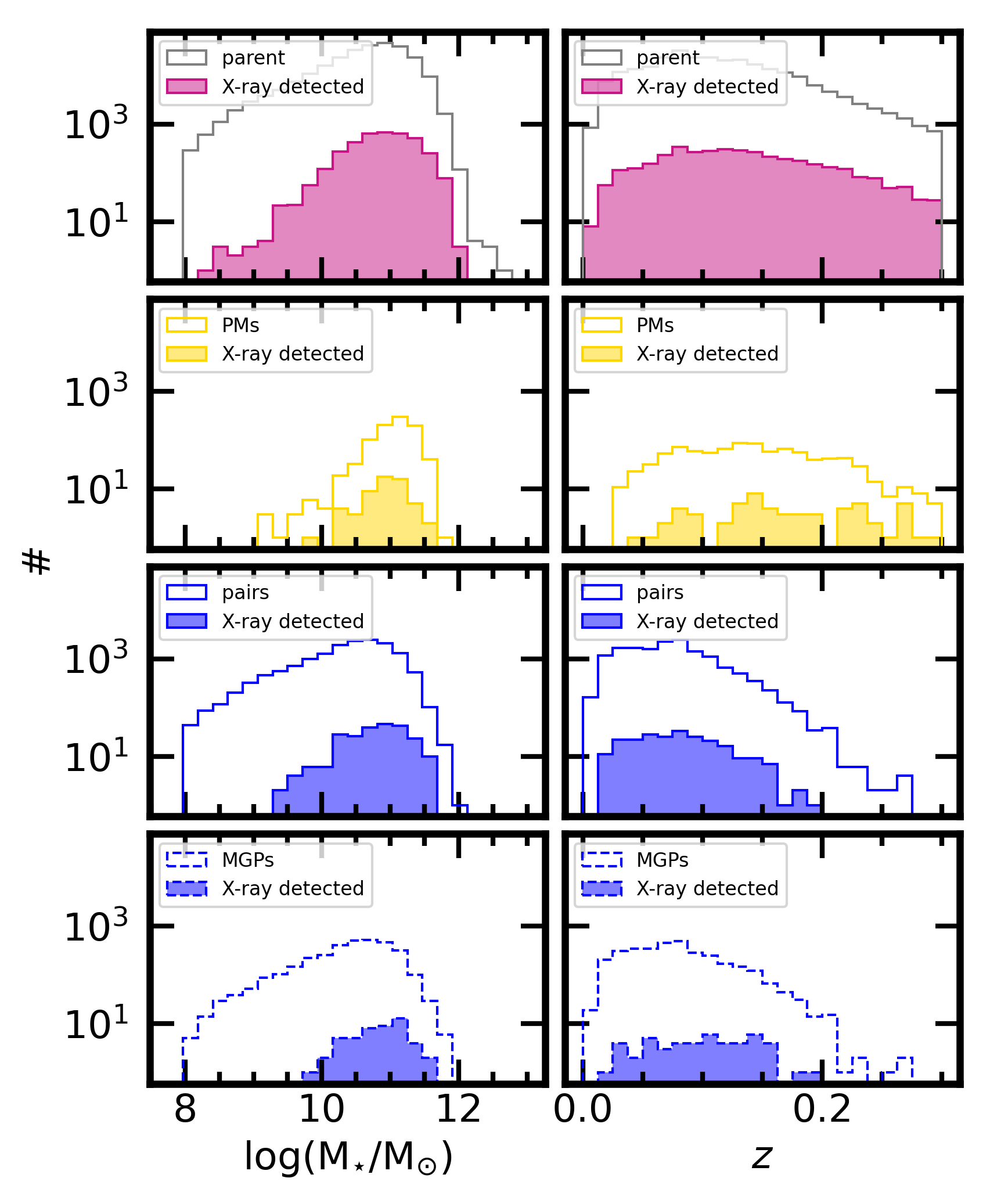}
\caption{Stellar masses (left column) and spectroscopic redshifts (right column) for galaxies in the parent sample (top row, grey histograms), further subdivided into X-ray AGNs from eRASS1 (top row, magenta histograms), post-mergers (second row, yellow), galaxies belonging to pairs meeting our criteria (third row, blue), and galaxies in mutual galaxy pairs (fourth row, blue dashed). For post-mergers, pairs, and MGPs, we also show the statistics for the X-ray detected subsets in the same panels (filled histogram series). Post-mergers and X-ray AGNs both lie at preferentially high stellar masses and redshifts due to the volume-limited nature of SDSS at high luminosity. Pairs and MGPs, meanwhile, are typically found at lower $z$ since fibre collision is less common, and galaxies in close pairs can be more reliably distinguished from one another.}
\label{fig:mz_stats}
\end{figure}

Figure~\ref{fig:mz_stats} shows the stellar mass (left panels) and redshift (right panels) distributions of the parent catalogue (grey series), the X-ray AGN sample (magenta), post-merger sample (yellow), pair galaxy sample (blue), and MGP sample (blue dashed). We also show the statistics for the sample of post-mergers, pairs, and MGPs that host X-ray detections in the filled histograms appearing in each panel. Relative to the parent sample, galaxy pairs tend to appear more often at lower $z$. This is due to the fibre collision effect of spectroscopic targets in SDSS DR7: galaxy pairs at close physical separations are more widely separated on the sky at low $z$, and the participant galaxies are more likely to be targeted individually for spectroscopy since the SDSS DR7 fibres are separated by a minimum of 55 arcsec (\citealp{2000AJ....120.1579Y}; \citealp{2008ApJ...685..235P}; \citealp{2016MNRAS.461.2589P}). The stellar mass distribution of pair galaxies traces the parent catalogue well. Meanwhile, the eRASS1 and post-merger samples lie at relatively high $z$, and $\mathrm{M_{\star}}$, compared to the parent sample. In both cases, this has to do with the volume-limited nature of the relevant survey at high luminosities. The morphological features associated with a major merger event are likely to be brighter on the sky when the galaxy is more intrinsically luminous, galaxies with high stellar masses are more likely to be identified as post-mergers, and we need to look out to higher $z$ to find them (the same effect is explored in the Appendix of \citealp{2023MNRAS.519.6149B}). An analogous selection bias results in the preferential appearance of highly X-ray luminous sources in eRASS1 at higher $z$.

\begin{figure}
\includegraphics[width=\columnwidth]{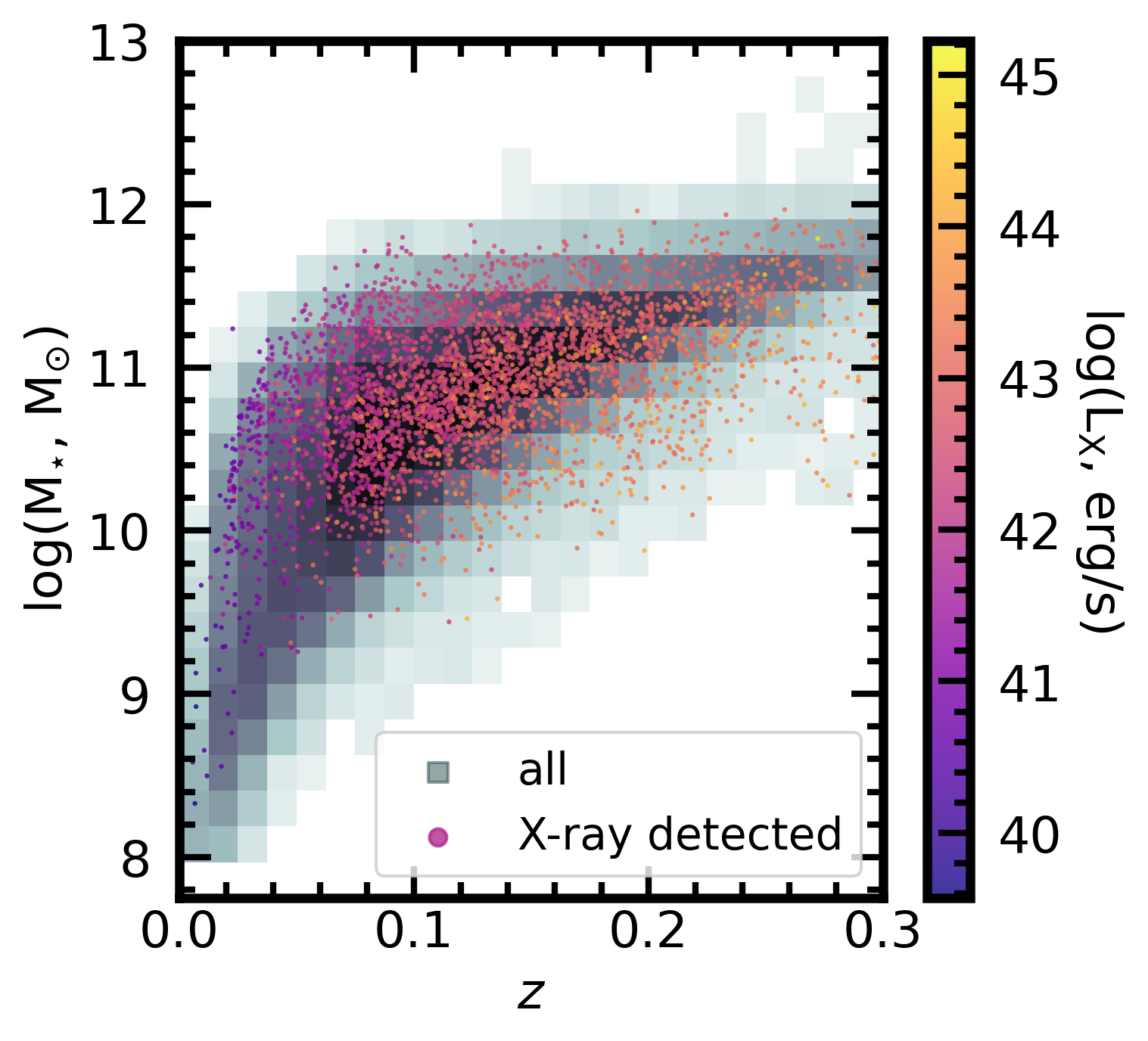}
\caption{The stellar mass-redshift distribution of the parent sample for this study (2D histogram), with eRASS1 detections superimposed. The colour scale represents X-ray luminosity in eRASS1.}
\label{fig:mzlx}
\end{figure}

Figure~\ref{fig:mzlx} presents $\mathrm{M_{\star}}$ and $z$ statistics for the parent catalogue (background histogram) and for eRASS1 sources (data points colour-coded by $\mathrm{L_{X}}$). The relative positions of the parent catalogue and eRASS1 sample on these axes highlights the difference between the SDSS DR7 and eRASS1 selection functions: at high redshifts, proportionally more X-ray AGNs are found in galaxies with intermediate masses, log($\mathrm{M_{\star}/M_{\odot}}$)\textasciitilde$10-11$. That the populations do not trace one another perfectly suggests that X-ray AGNs in eRASS1 are preferentially hosted by galaxies in this mass domain. Still, in Section~\ref{Results}, all results account for such biases by performing rigorous matching between merger and non-interacting control samples on $\mathrm{M_{\star}}$ and $z$.

\subsubsection{Pairwise treatment of X-ray detections in galaxy pairs}
\label{Pairwise treatment of X-ray detections in galaxy pairs}

When assessing X-ray AGN incidence in galaxy pairs, we expect that close pairs are likely to appear as crowded within the \textasciitilde30 arcsec eROSITA PSF (which corresponds to a projected physical distance of \textasciitilde57 kpc at $z=0.103$, the median redshift of the parent catalogue studied in this work). We must therefore count X-ray AGN status on a pairwise basis, rather than considering the properties of individual galaxies. We study the sample of MGPs whenever X-ray incidence is considered in the context of galaxy pairs. If either or both MGP galaxies have been assigned to an X-ray source in the eRASS1 catalogue, we count the MGP collectively as an AGN. If neither has been assigned to an X-ray source, we count it as a non-AGN. The same approach is used for the controls; although separate controls are matched to each member of the MGP, we count the AGNs pairwise in their controls.  In this way, we are fairly counting X-ray detections between the MGPs and their controls.

Given our pairwise approach to AGN statistics, it is necessary to derive a statistical correction for the influence of X-ray PSF blending on the sample of MGPs. The measured AGN fraction amongst MGPs in a given bin of projected separation ($f_{m}$) is a function of the true AGN fraction ($f_{A}$) and the fraction of galaxies crowded below the effective resolution of the survey ($f_{c}$). We can write $f_{A}$ in terms of $p_{0}$, the chance of a galaxy pair hosting zero AGNs, $p_{s}$, the chance of a pair with a single AGN, and $p_{d}$, the chance of a pair with a double AGN. The coefficient for $p_{d}$ is one, since a MGP with two AGNs has an AGN fraction of 1, and the coefficient for $p_{s}$ is $\frac{1}{2}$, since a MGP with only a single AGN has an AGN fraction of one half. Since a given pair must have either zero, one, or two AGNs, $p_{0}$, $p_{s}$ and $p_{d}$ sum to one, allowing a simplification:

\begin{equation}
	f_{A}=\frac{p_{d}+\frac{1}{2}p_{s}}{p_{0}+p_{d}+p_{s}}=p_{d}+\frac{1}{2}p_{s}.
	\label{eq:fa_def}
\end{equation}

When crowding occurs at a rate $f_{c}$, the measured AGN fraction $f_{m}$ is no longer the same as the actual AGN fraction $f_{A}$. In cases where double-AGN pairs are blended within the PSF, $p_{d}$ is multiplied by a crowding factor ($1-\frac{1}{2}f_{c}$), since only half of the AGNs are counted in a crowded MGP with a double AGN. Meanwhile, there is no effect on the number of AGNs detected in single-AGN pairs, so the coefficient for $p_{s}$ remains the same at $\frac{1}{2}$:

\begin{equation}
	f_{m}=(1-\frac{1}{2}f_{c})p_{d}+\frac{1}{2}p_{s}.
	\label{eq:fm_def}
\end{equation}

In order to correct the measured AGN fraction up to the true AGN fraction, we are interested in $f_{m}/f_{A}$, which can be written as the ratio of Equations~\ref{eq:fa_def} and~\ref{eq:fm_def}:

\begin{equation}
	\frac{f_{m}}{f_{A}}=\frac{(1-\frac{1}{2}f_{c})p_{d}+\frac{1}{2}p_{s}}{p_{d}+\frac{1}{2}p_{s}}.
	\label{eq:fmoverfa}
\end{equation}

Conveniently, $p_{d}$, $p_{0}$, and $p_{s}$ can all be written in terms of $f_{A}$. The likelihood of an MGP with two AGNs is set by combining the independent AGN likelihoods of two galaxies:

\begin{equation}
	p_{d}=f_{A}^{2}.
	\label{eq:sub1}
\end{equation}

The likelihood of an AGN-free MGP is also the combination of the independent chances that two galaxies do not host an AGN:

 \begin{equation}
	p_{0}=(1-f_{A})^{2}.
	\label{eq:sub3}
\end{equation}

Any MGPs that are neither double AGNs nor AGN-free must be single AGNs, represented by $p_{s}$. We can therefore subtract $p_{d}$ and $p_{0}$ from the total to obtain $p_{s}$:

\begin{equation}
	p_{s}=1-f_{A}^{2}-(1-f_{A})^{2}=2(f_{A}-f_{A}^{2}).
	\label{eq:sub2}
\end{equation}

Substituting the last three definitions into Equation~\ref{eq:fmoverfa}, we can rewrite and simplify (with some intermediate steps omitted for brevity):

\begin{equation}
	\frac{f_{m}}{f_{A}}=\frac{-1}{2}f_{A}f_{c}+1.
	\label{eq:ratio}
\end{equation}

Finally, we derive the expression for the true AGN fraction:

\begin{equation}
	f_{A}=\frac{1-\sqrt{1-2f_{c}f_{m}}}{f_{c}}.
	\label{fin}
\end{equation}

In combining the AGN statuses of MGPs and control pairs, we force a crowding fraction of unity, $f_{c}$=1. This avoids the difficulty of attempting to predict which galaxy pairs in our sample will be crowded. With $f_{c}$ methodologically forced to 1, the true AGN fraction is $1-\sqrt{1-2f_{m}}$. This expression yields in the two extreme cases: when there are no AGNs ($f_{m}=0$ and $f_{A}=0$), and when all galaxies in a sample host an AGN ($f_{m}=0.5$ and $f_{A}=1$, since both single- and double-AGN pairs would be detected as single AGNs). This method and correction are applied to both the MGPs and matched control samples when we study X-ray AGN incidence in galaxy pairs.

\subsubsection{BLAGN sample}
\label{BLAGN sample}

Galaxies in this work are counted as broad-line AGNs (BLAGNs) if they are included in the sample derived homogeneously from SDSS DR7 spectroscopic targets by \citet{2019ApJS..243...21L}. \citet{2019ApJS..243...21L} use a sophisticated spectral fitting technique that allows for multiple components to the Balmer and forbidden lines appearing in the spectra, and include galaxies in their final BLAGN sample if the quality of the fit is improved by a broad H$\alpha$ component, and if that component is detected with S/N>5. There is no lower limit on the velocity of the broad H$\alpha$ component, so long as it is broader than the narrow component. In practice, BLAGNs in the \citet{2019ApJS..243...21L} catalogue have H$\alpha$ line velocities of at least 428 km/s, and a median velocity of 2,763 km/s. In our parent catalogue, 2,626 galaxies are identified as BLR hosts by \citet{2019ApJS..243...21L}. For additional detail, we refer the reader to Tables 1 and 2 of \citet{2019ApJS..243...21L}.

\subsubsection{NLAGN sample}
\label{NLAGN sample}

In order to identify galaxies with narrow-line AGN (NLAGN) emission lines, we refer to the widely-used MPA-JHU catalogue of spectroscopic measurements for SDSS DR7 (\citealp{2004MNRAS.351.1151B}). While the emission line fitting method used by MPA-JHU considers only one component each for Balmer and forbidden emission lines, it is well suited to characterize emission line fluxes in the absence of a dominant BLR contribution to the optical spectrum. MPA-JHU also derive their measurements from the SDSS DR7 spectroscopic sample, and so the entire parent catalogue is technically eligible to be classified by the MPA-JHU emission line measurements as a NLAGN (even though emission line measurements for every source in the parent catalogue are not tabulated). MPA-JHU apply the following criteria to SDSS DR7 sources before reporting emission lines used in this work to select NLAGNs:

\begin{itemize}
   \item Fluxes are always reported when the spectroscopic class and the target type are "galaxy", the redshift is less than 0.7, and the spectrum has a median pixel-wise S/N>0.
   \medskip
   \item Fluxes are also reported when the spectroscopic class is "galaxy" (even when the target type is not "galaxy"), when the redshift is less than 0.7, and the pixel-wise S/N>2.
   \medskip
   \item Fluxes are reported for objects with spectroscopic class "quasar" when the object was targeted as a galaxy, the redshift is less than 0.7, the S/N is >2, the $\chi^{2}$<2 for the spectral fit to the continuum, the (single-component) Balmer line width is <500 km/s, and the forbidden lines are no less than 85\% as wide as the Balmer lines.
\end{itemize}

Given the velocity thresholds in the MPA-JHU catalogue, and the typical velocities in the \citet{2019ApJS..243...21L} BLAGN sample, the two catalogues are generally mutually exclusive. In the 469 (28\% of BLAGNs) cases where a galaxy exists in both the MPA-JHU and \citet{2019ApJS..243...21L} catalogues, we prioritize the \citet{2019ApJS..243...21L} classification as a BLAGN. For the remaining (narrow emission line) galaxies in the MPA-JHU catalogue, we assess the presence of an AGN using the \citet{2001ApJ...556..121K} criterion on the BPT diagram, requiring also that each of the four lines used for BPT classification are detected with S/N>5, and that the H$\alpha$ equivalent width is greater than 6\AA. The first criterion is designed to select galaxies with narrow line nebular emission unambiguously dominated by an AGN contribution, while the second is deployed to remove galaxies dominated by low-ionization emission regions (LIERs, e.g., \citealp{2006MNRAS.371..972S}; \citealp{2011MNRAS.413.1687C}; \citealp{2016MNRAS.461.3111B};  typically produced by shocks, weak AGNs, or old / hot stellar populations) from the sample. Removal of LIERs is important in merger studies, since hydrodynamic shocks are an intuitive (and studied; see \citealp{2006ApJ...637..138M,2010ApJ...724..267F,2014ApJ...781L..12R,2015ApJS..221...28R}) consequence of galaxy mergers. While the choice of BPT diagram criteria (e.g., a lower S/N criterion, including composite galaxies between the \citealp{2001ApJ...556..121K} and \citealp{2003MNRAS.346.1055K} criteria on the diagram) does bear on the NLAGN fraction in the sample and some of our results, it does not affect the trends presented in this work. There are 2,723 galaxies in the parent catalogue that meet our criteria for classification as a NLAGN.

\subsubsection{Mid-IR AGN sample}
\label{Mid-IR AGN sample}

Characterization of dust-obscured AGNs is central to this work, and we use photometry from the W1 and W2 bands of the WISE space telescope to select galaxies whose circum-nuclear or galaxy-scale dust has been heated by UV emission from the accretion disk. Depending on the geometry of the dust, mid-IR observations can be used to identify galaxies whose X-ray emission, BLR, and/or NLR are not observationally detected. Since our sample lies at low-$z$, we follow several literature efforts (e.g., \citealp{2014MNRAS.441.1297S}; \citealp{2013MNRAS.435.3627E}) and use a single colour criterion, in this case W1-W2>0.5, to select galaxies with a dust-obscured AGN. In order to ensure that our mid-IR data products are homogeneously available (and available with high S/N) for the entire SDSS DR7 parent catalogue described in Section~\ref{Parent catalogue}, we use W1 and W2 Vega magnitudes derived from unWISE forced photometry\footnote{Forced photometry refers to de-blended photometry from WISE observations at the locations of galaxies in SDSS.} at the positions of SDSS galaxies (\citealp{Lang_2016}). As a result, no sources from our parent catalogue are missing W1 or W2 photometry, and we can characterize the hot dust in our entire galaxy sample. In the parent catalogue, 6,391 galaxies are classified as mid-IR AGNs using these criteria.

\subsubsection{Multi-wavelength AGNs}
\label{Multi-wavelength AGNs}

Having laid out the criteria used to identify AGNs in the X-ray, optical (narrow- and broad-line diagnostics) and mid-IR, we can investigate the global connections between the different AGN classes independent of the merger sequence.

\begin{figure*}
\includegraphics[width=\textwidth]{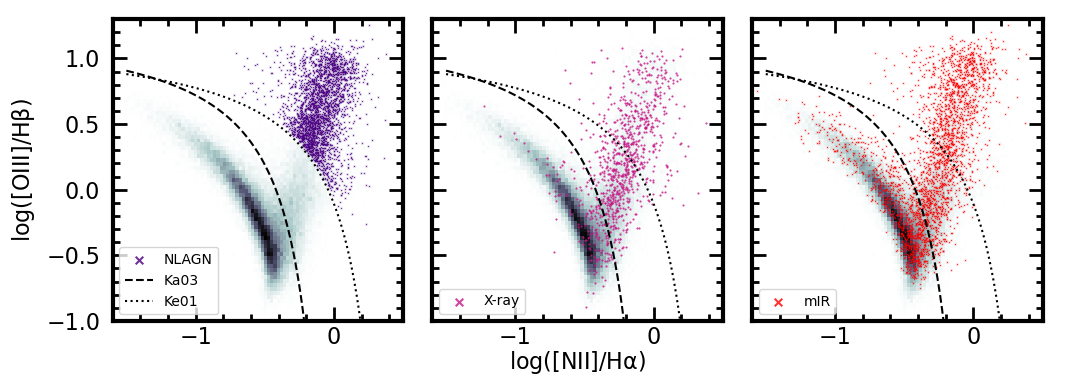}
\caption{BPT AGN positions of the entire sample (2D histograms in each panel), galaxies meeting our NLAGN criteria (indigo points, left panel), galaxies with X-ray detections eligible for placement on the BPT diagram (middle) and galaxies with unWISE W1-W2 colours suggesting a dust-obscured AGN (right) eligible for placement on the BPT diagram.}
\label{fig:bpt_stats}
\end{figure*}

Figure~\ref{fig:bpt_stats} shows the BPT diagram positions of the parent catalogue (background histogram in each panel), NLAGNs (left panel, indigo), X-ray AGNs (middle panel), and mid-IR AGNs (right panel). Notably, galaxies meeting our other AGN criteria do not always lie in the AGN wing of the BPT diagram; the positions of the X-ray AGNs echo the results from \citet{2019ApJ...876...12A} (and other works investigating the incompleteness of NLAGN selection, e.g., \citealp{2015ApJ...811...26T}; \citealp{2016ApJ...826...12J}). Mid-IR AGNs are even more widely distributed throughout the BPT diagram with 39\% of BPT-eligible (with S/N>5 for each of the four BPT diagram optical emission lines) mid-IR AGNs falling in the star-forming locus of the diagram defined by the \citet{2003MNRAS.346.1055K} criterion (dashed lines, Figure~\ref{fig:bpt_stats}). Taken as a whole, the narrow-line characteristics of the various AGN classes highlight the necessity of multiple AGN criteria for a complete study. We do not show the positions of BLAGNs in Figure~\ref{fig:bpt_stats} because they are de facto absent from the MPA-JHU catalogue.

\begin{figure*}
\includegraphics[width=\textwidth]{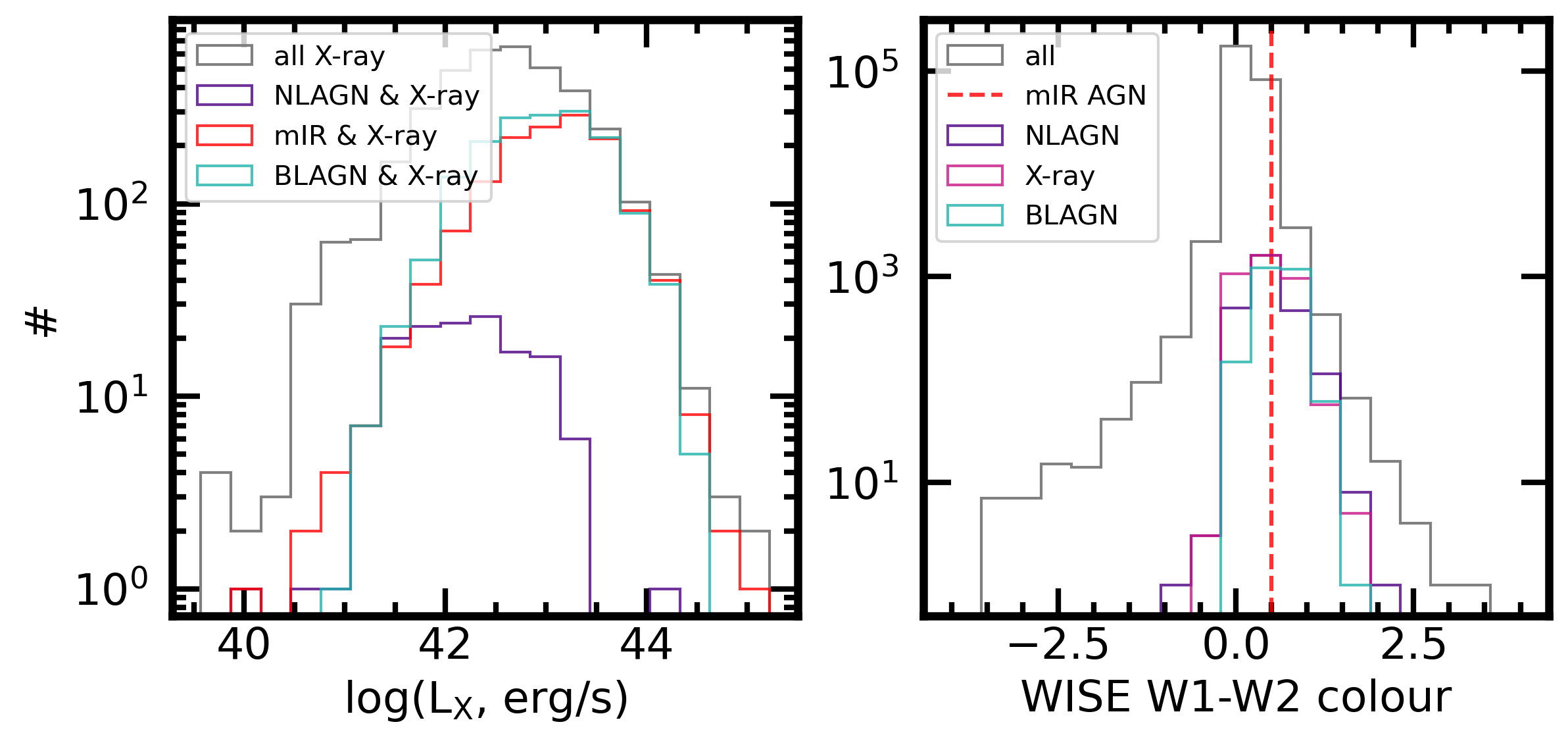}
\caption{Observed X-ray luminosities from eRASS1 (left) and W1-W2 colour from unWISE (right) for the galaxies eligible for this study (grey series). The galaxies with X-ray detections (left) that also meet our narrow-line (indigo), broad-line (cyan), or mid-IR (red) AGN criteria are also plotted. The W1-W2 Vega magnitude colours from unWISE (right) are also shown for the subsets of galaxies meeting our other AGN criteria (same colour code as left, and with eRASS1 sources shown in magenta).}
\label{fig:mw_stats}
\end{figure*}

Figure~\ref{fig:mw_stats} shows the observed X-ray luminosities (left panel) and mid-IR colours (right panel) of galaxies meeting the various AGN criteria in this work. Generally, galaxies with NLAGN classifications lie at lower X-ray luminosities than those with either BLAGN or mid-IR AGN detections. The dominance of BLAGNs and mid-IR AGNs at high $\mathrm{L_{X}}$ hints at the accretion rates corresponding to each AGN type, although the effect is partly caused by BLAGNs "forcing out" NLAGNs from the high $\mathrm{L_{X}}$ domain, since BLAGN classifications supersede NLAGN classifications in this work. NLAGNs with low observed X-ray luminosity may also be attenuated along the line of sight; in such cases the observed luminosity would not be representative of the intrinsic state of the AGN. Evidence for lower rates of X-ray obscuration amongst BLAGNs is presented later in Section~\ref{Multi-wavelength observability of X-ray AGNs in mergers}. The WISE colours of the galaxies in the sample illustrate that galaxies with other AGN classifications do not always meet our mid-IR AGN criterion (to the right of the red dashed line), a result expected in cases where the AGN is not bolometrically dominant.

\begin{figure}
\includegraphics[width=\columnwidth]{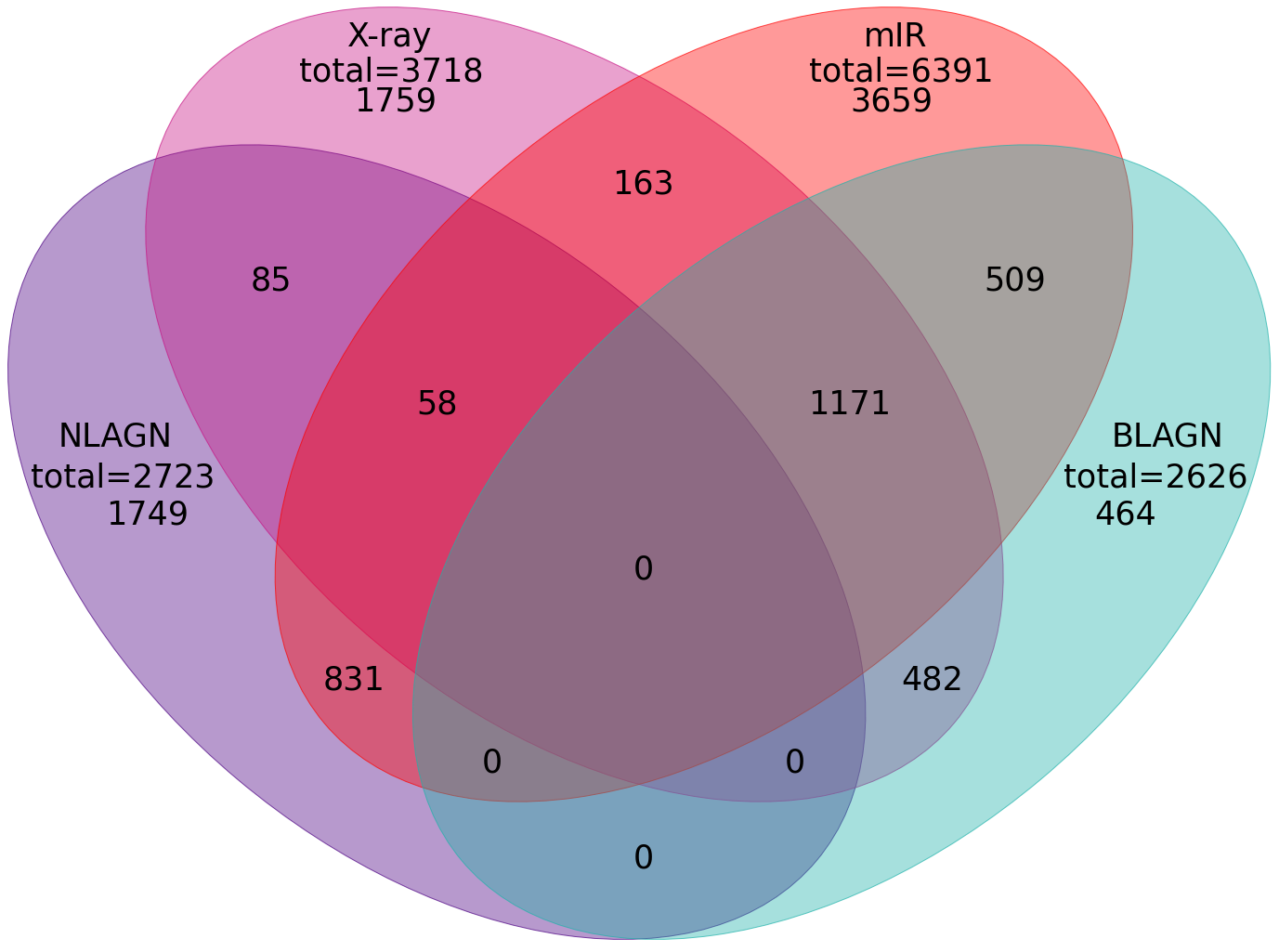}
\caption{Four-class Venn diagram describing the overlap between the AGN criteria in our sample. Overlap exists between all classes, except where it is expressly forbidden (our BPT AGN criteria, indigo, and BLAGN criteria, cyan, are explicitly mutually exclusive).}
\label{fig:agn_venn}
\end{figure}

Figure~\ref{fig:agn_venn} summarizes the multi-wavelength AGN selection used in this work. Overlap exists between all classes in the sample, except where it is forbidden by construction (BLR detection forbids the inclusion of a galaxy in the NLAGN sample). While this work prioritizes the novel statistical results made possible by the eRASS1 X-ray catalogue, the intersection of the X-ray AGN sample with other AGN types offers essential context. It is worth noting that our BLAGN (Section~\ref{BLAGN sample}), NLAGN (Section~\ref{NLAGN sample}), and mid-IR (Section~\ref{Mid-IR AGN sample}) samples are selected homogeneously from the SDSS DR7 catalogue, thus ensuring that any object in our parent catalogue could be classified as an AGN by any of our multi-wavelength AGN criteria. The most striking connection on the Venn diagram is the intersection between the X-ray, mid-IR, and BLAGN samples (overlap of the magenta, red, and cyan ovals). In particular, the overlap between the X-ray and BLAGN ovals emphasizes the importance of including BLAGNs in our analysis; without them, \textasciitilde44\% of the X-ray AGN sample would be eliminated. Figure~\ref{fig:agn_venn} also illustrates the utility of X-ray selection in multi-wavelength AGN studies: nearly half (47\%) of the X-ray AGNs (the magenta oval) are not identified as AGNs according to any other criteria.

\begin{figure}
\includegraphics[width=\columnwidth]{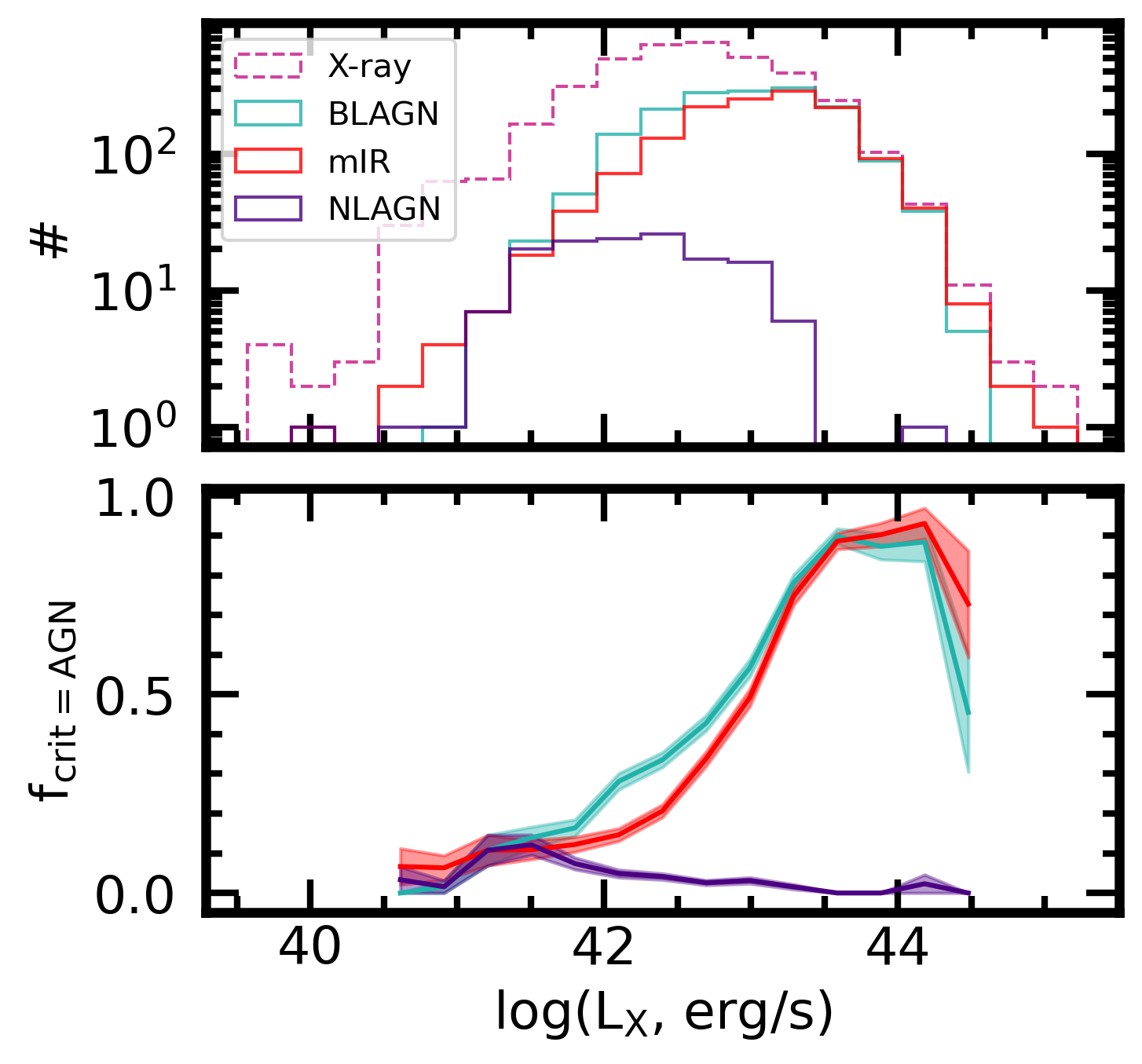}
\caption{The number (top panel) and fraction (bottom panel) of eRASS1 galaxies in a given bin of observed X-ray luminosity meeting one of our other AGN criteria (NLAGN, mid-IR, or BLAGN). Fractions are only plotted when there are more than 10 X-ray galaxies in a given bin of $\mathrm{L_{X}}$. The shaded error regions are defined by the binomial error on the fraction, $\sqrt{f(1-f)/N}$ where $f$ is the measured fraction and $N$ is the number of X-ray galaxies in the bin.}
\label{fig:agn_fracs}
\end{figure}

Except in cases where obscuration is significant, ($\mathrm{N_{H}}\gtrsim10^{22}$/cm$^{2}$) X-ray luminosity is a comparatively reliable indicator of the state of the SMBH, since the BLR is more easily obscured by dust, and the observability of NLAGNs and mid-IR AGNs is also conditional. Figure~\ref{fig:agn_fracs} characterizes the observed X-ray luminosities most often associated with each of our other AGN criteria. The top panel (repeated for context from the left panel of Figure~\ref{fig:mw_stats}) shows the luminosity distributions of the entire X-ray AGN sample (magenta dashed histogram), as well as the subsets of the sample meeting our BLAGN (cyan), mid-IR (red), and NLAGN (indigo) criteria. The bottom panel shows the fractions of X-ray AGNs co-detected as BLAGNs, mid-IR AGNs, and NLAGNs as a function of $\mathrm{L_{X}}$. X-ray AGNs are most often co-detected as NLAGNs at the low end of luminosities in our sample, between \textasciitilde$10^{41-42}$ erg/s (shown also in Figure~\ref{fig:mw_stats}).

The fraction(s) of X-ray galaxies co-detected as BLAGNs and/or mid-IR AGNs grows steadily with increasing $\mathrm{L_{X}}$, and there is nearly total overlap between the X-ray, BLAGN, and mid-IR samples for $\mathrm{L_{X}}$>$10^{43.5}$ erg/s. The multi-wavelength demographics of the X-ray sample suggest that hot circum-nuclear or galaxy-scale dust is more often observed in galaxies with higher observed AGN luminosities. Higher X-ray luminosities are also frequently associated with the detection of an exposed BLR, consistent with a higher likelihood of removal of obscuring material (dust and/or gas; \citealp{2003ApJ...598..886U,2007A&A...468..979M,2008ApJ...679..140T,2011ApJ...728...58B,2017Natur.549..488R}) or degrees of attenuation of the X-ray flux by the obscuring material.

Implicitly absent from Figure~\ref{fig:agn_fracs} are low-luminosity or obscured AGNs without an X-ray detection (fainter than the eRASS1 flux limit at $5\times10^{-14}$erg/s/cm$^{-2}$ in the 0.2$-$5 keV band). Still, the general AGN demographics of the sample suggest an AGN "observability sequence" with $\mathrm{L_{X}}$, in which AGNs with low intrinsic X-ray luminosities, heavy obscuration, or degrees of both are often seen as NLR hosts. Progressively higher $\mathrm{L_{X}}$ is consistent with being concurrent with removal of the NLR, exposure of the BLR, and more prominent nuclear hot dust emission relative to the contribution of starlight to the mid-IR (\citealp{2017Natur.549..488R}; \citealp{2018MNRAS.478.3056B}). Understanding these trends and the interplay between multi-wavelength diagnostics will be vital in our interpretation of our merger results, presented in the next section.

\section{Results}
\label{Results}

After assembling a catalogue with homogeneous multi-wavelength data, we commence our study of the influence of pre- and post-coalescence galaxy mergers on the X-ray and multi-wavelength AGNs in the sample.

\subsection{X-ray AGN incidence in mergers}
\label{X-ray AGN incidence in mergers}

\begin{figure}
\includegraphics[width=\columnwidth]{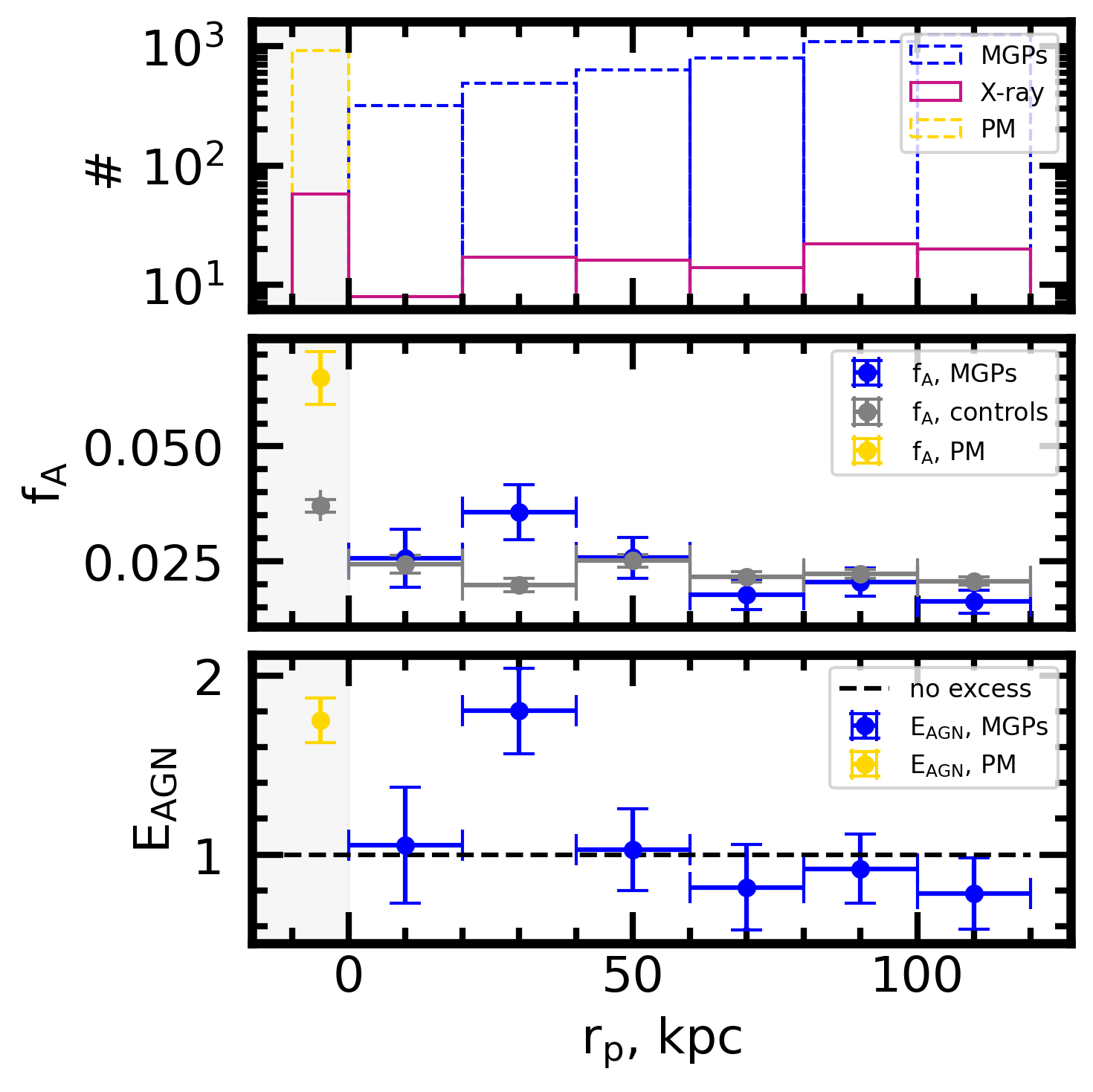}
\caption{The number (top), fraction (middle), and excess over matched controls (bottom) of X-ray detections in our sample as a function of projected separation, using our pairwise treatment for galaxy pairs. Error bars on the AGN fractions are the binomial errors, and the errors on the excess are calculated as $\mathrm{\sigma_{f_{Mer}}/f_{Mer}+\sigma_{f_{Ctrl}}/f_{Ctrl}}$ where $\mathrm{f_{Mer}}$ is the AGN fraction in the merger sample and $\mathrm{f_{Ctrl}}$ is the AGN fraction in the controls.}
\label{fig:xagn_exc}
\end{figure}

Numerous statistical studies (e.g., \citealp{2011MNRAS.418.2043E,2013MNRAS.435.3627E}; \citealp{2014MNRAS.441.1297S}; \citealp{10.1093/mnras/stw2620}; \citealp{2023MNRAS.519.6149B}) have used optical and mid-IR diagnostics to uncover an increasing AGN fraction in mergers at progressively closer separations, with a peak in the post-merger regime. Although some previous studies have assessed X-ray AGN frequency in low-$z$ galaxy mergers, statistics have either been poor (e.g., \citealp{2020MNRAS.499.2380S}) and/or limited in the merger stage that they assess (e.g., \citealp{2023MNRAS.523..720L}). Thanks to our eRASS1 sample, we are now able to make the same assessment for X-ray selected AGNs in a large statistical sample of both galaxy pairs and post-mergers for the first time.

As mentioned earlier, the blending of pairs of galaxies within the eROSITA PSF means that we treat pre-coalescence galaxies in a pairwise fashion. Hence, for this analysis, we will use the MGP sample. MGPs and their matched controls are sorted into six 20 kpc bins (from $0-120$ kpc). Their X-ray AGN fractions are assessed and corrected using the method described in Section~\ref{Pairwise treatment of X-ray detections in galaxy pairs}. For post-mergers, we match individual galaxies to individual controls, since post-mergers do not belong to crowded galaxy pairs. The mergers in each subset all succeed in finding 10 matched isolated control galaxies each, with no significant discrepancy in $\mathrm{M_{\star}}$ or $z$ statistics. In the pair, post-merger, and control samples, we count AGN detections and fractions, and then divide the AGN fractions in each merger sample by the AGN fractions in the relevant control sample to compute an excess.

The results of this experiment are shown in Figure~\ref{fig:xagn_exc}, which shows the numbers of MGPs in each bin of $\mathrm{r_{p}}$ and number of post-mergers (blue and yellow histograms, respectively) with the number of X-ray detections associated with each bin in the top panel. The middle panel shows the recovered AGN fractions in the merger samples and matched control samples. Controls have fairly consistent $f_{A}$ of \textasciitilde$2-2.5$\%, although the control sample for post-mergers is slightly higher (\textasciitilde3.5\%) because the post-mergers and their controls are offset towards higher masses (see Figure~\ref{fig:mz_stats}). The bottom panel shows the excesses computed by taking the ratio of $f_{A}$ in mergers and controls. In line with previous studies of optical and mid-IR selected AGNs in galaxy pairs (e.g., \citealp{2013MNRAS.435.3627E}; \citealp{2014MNRAS.441.1297S}), pairs at wide separations ($\mathrm{r_{p}}$>40) have X-ray AGN fractions that are statistically similar to non-interacting galaxies (although, the statistical significance, strength, and physical extent in $\mathrm{r_{p}}$ of the excess varies with sample size and methodology), suggesting that the influence of early pair-phase interactions are ineffective at triggering AGNs with observed luminosities detectable in eRASS1 at these separations. Pairs with $\mathrm{r_{p}}$ between $20-40$ kpc do appear to host X-ray AGNs in excess of the isolated control sample by a factor of \textasciitilde1.8. The excess indicates that the merger sequence has begun to drive gas inflows strong enough to trigger eRASS1-detectable AGNs at these separations. Optical and mid-IR AGN fractions in \citet{2023MNRAS.519.6149B} first become statistically distinguishable from controls near \textasciitilde40 kpc, even though the method is not identical.

Perhaps unexpectedly, the closest galaxy pairs with $\mathrm{r_{p}}$<20 kpc in our sample do not appear to host X-ray AGNs in excess of their matched controls. The lack of an AGN excess in close galaxy pairs is in contrast to the findings in \citet{2023MNRAS.519.6149B} and at other wavelengths in this work (see Section~\ref{Multi-wavelength AGN excesses}). It is unlikely that X-ray AGNs triggered by mergers have actually "switched off" at these separations. Rather, we hypothesize that merger-induced obscuration is effectively masking the underlying AGN excess in this bin of $\mathrm{r_{p}}$, and the intrinsic AGN luminosities triggered by pair-phase interactions are not sufficient to remove the obscuring material via radiation pressure. We offer observational evidence for this hypothesis later in Section~\ref{Obscuration excess}.

The X-ray AGN excess recovered for post-mergers is once again in agreement with previous results: we find that X-ray AGNs appear to be triggered \textasciitilde1.8$\pm$0.1 times as frequently in post-mergers compared to isolated controls, within the error bars of \citet{2020MNRAS.499.2380S} and \citet{2023ApJ...944..168L}; we return to a more detailed comparison with the literature in Section~\ref{Discussion}. The positive and significant excess of X-ray AGNs in post-mergers verifies the role of coalescence in triggering luminous AGNs.

\subsection{Post-mergers in X-ray AGNs}
\label{Post-mergers in X-ray AGNs}

In Figure~\ref{fig:xagn_exc}, we demonstrated that there is an excess of X-ray AGNs in galaxy pairs with separations of $20-40$ kpc and in post-mergers, indicating that, statistically, mergers can trigger AGNs. We now turn to the complementary experiment (e.g., \citealp{2017MNRAS.470..755H}; \citealp{2019MNRAS.487.2491E}) and assess what fraction of AGNs are in mergers. This experiment aims to assess the prevalence of mergers as the AGN triggering mechanism. We use the same control matching approach to match the X-ray detected eRASS1 galaxy sample to a control pool of X-ray-undetected galaxies (i.e., non-AGNs). Because our X-ray galaxies are significantly offset towards higher masses, control matching is truncated after 3 batches when the $\mathrm{M_{\star}}$ K-S $p$-value falls just below our quality control criterion to 0.89. We then compute the fraction of X-ray AGNs and non-AGNs that have a post-merger classification. Of the 3,718 X-ray AGNs studied, $1.6\pm0.2$\% have a post-merger classification, and $0.8\pm0.08$\% of the 11,154 controls have a post-merger classification. We report a post-merger excess of $2.0\pm0.24$ in X-ray detected galaxies over X-ray undetected controls. The post-merger fractions of the X-ray detected and undetected control samples (1.6\% and 0.8\%, respectively) are lower limits, since our post-merger selection prioritizes purity over completeness. However, the quantitative post-merger excess remains accurate; since the post-merger sample was identified independently of the X-ray AGN sample, and the fractional post-merger completeness is in principle identical in both the X-ray and non-X-ray samples. Combining this evidence with our X-ray AGN excess result for post-mergers, we conclude that mergers and X-ray AGNs are unambiguously linked at low-$z$.

\subsection{X-ray luminosity in mergers}
\label{X-ray luminosity in mergers}

Having established that some pair-phase galaxy interactions and coalescence events both appear to trigger luminous X-ray AGNs, we next narrow our focus and compare the observed luminosities of AGNs in mergers to AGNs in non-merging isolated galaxies. The comparison of AGNs in mergers and AGNs in non-mergers again requires control matching. We use the same galaxy-wise control method described in Section~\ref{Control pool} to identify at least 5, and as many as 10 controls with an eRASS1 detection for each merger-AGN (in either our pair or post-merger sample, also requiring an eRASS1 detection). We use an ensemble of at least 5 controls per merger because the goal of this experiment is to determine whether the merger's X-ray luminosity is atypical for galaxies with similar $\mathrm{M_{\star}}$ and $z$. Every merger subset (post-mergers and bins of galaxy pairs) finds 10 controls per merger except for the 60<$\mathrm{r_{p}}$<80 kpc bin, which finds 5 batches before the K-S $p$-value for $z$ falls to 0.899. We finally calculate a log-scale X-ray luminosity offset $\mathrm{\Delta log(L_{X})}$ for each merger by subtracting the mean log X-ray luminosity of the control ensemble from that of the merger.

\begin{figure}
\includegraphics[width=\columnwidth]{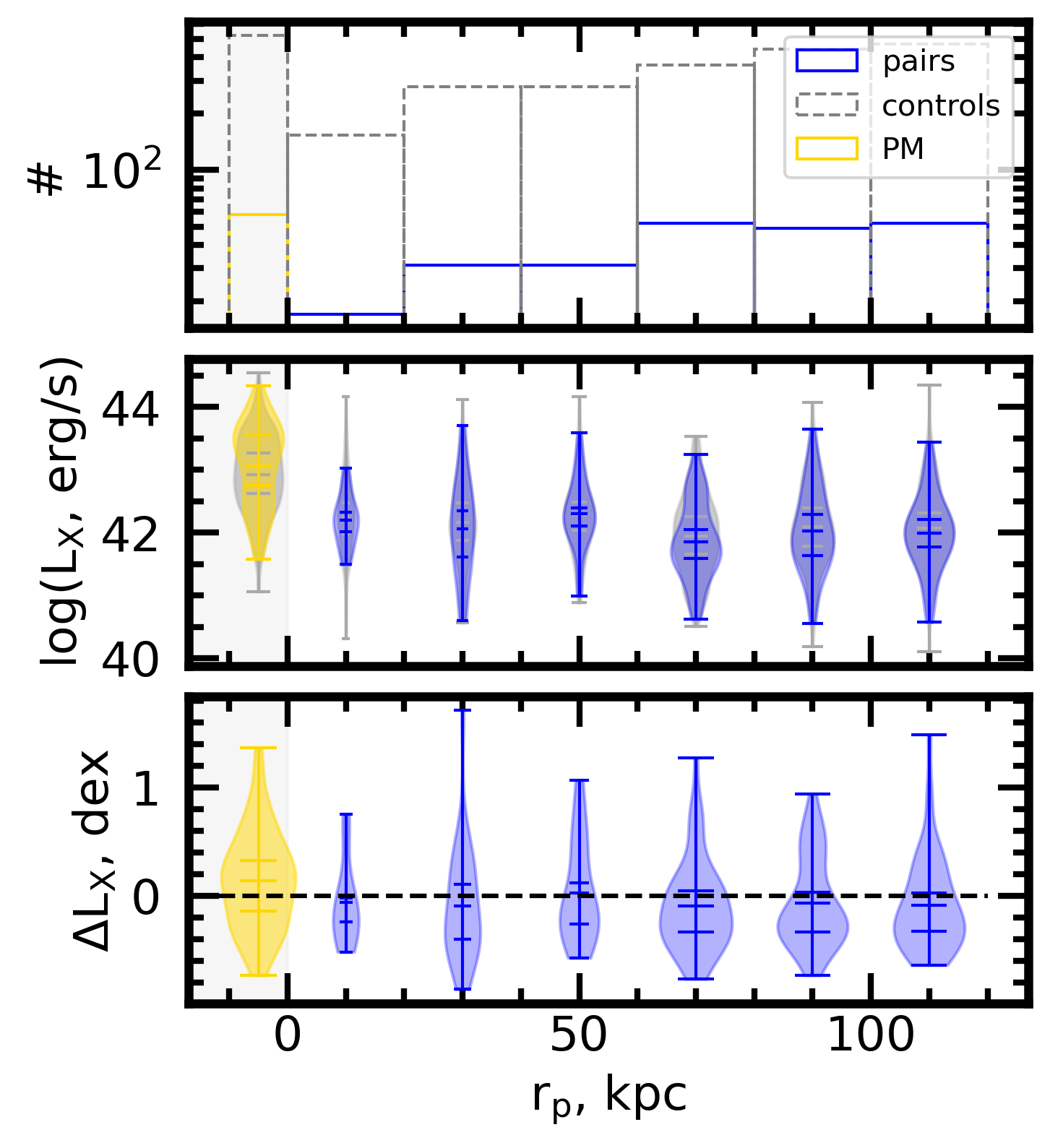}
\caption{Observed X-ray luminosity offset plot for galaxy pairs (blue) and post-mergers (yellow) in our sample relative to X-ray AGNs in isolated controls (grey). The top panel shows the numbers of post-merger and pair AGNs, and the number of control AGNs matched for each merger sample. The middle panel violin plot shows the luminosity distributions for the merger and control samples, with tickmarks at the extrema, means, and $1\sigma$ positions. The bottom panel violin series shows the distribution of luminosity enhancements calculated for the mergers compared to their individual control ensembles. X-ray AGNs in pairs are shown in the blue data series, and post-mergers are shown in yellow. Galaxy pairs and post-mergers are both found to have uncorrected X-ray luminosities in eROSITA that are statistically consistent with their non-interacting controls. The violin series in the third panel highlights that galaxies across the merger sequence exhibit a range of X-ray luminosities.}
\label{fig:lum_enh}
\end{figure}

Figure~\ref{fig:lum_enh} shows the results of this test. The top panel histogram shows the numbers of pair galaxies in each bin of $\mathrm{r_{p}}$ (blue) and post-mergers (yellow), as well as the number of isolated controls matched to them (grey). Note that all post-mergers, pairs, and controls included in this experiment must have an eRASS1 detection associated with them, so that $\mathrm{\Delta log(L_{X})}$ can be computed. The middle panel violin plots show the $\mathrm{L_{X}}$ distributions of the merger samples (in either blue or yellow for pairs and post-mergers, respectively) and the control galaxies matched to them (grey). The tick marks on the violins show the maxima and minima of each $\mathrm{log(L_{X})}$ distribution, the means, and the $1\sigma$ region. Finally, the bottom panel shows the actual result of our experiment, in which the median log($\mathrm{L_{X}}$) of the control ensemble for each merger is subtracted from the merger's $\mathrm{log(L_{X})}$ to compute a luminosity enhancement. The distribution of $\mathrm{\Delta log(L_{X})}$ values for each merger sample is used to create the violin series.

The distributions of $\mathrm{log(L_{X})}$ for mergers and their matched controls are relatively consistent (reasonable overlap between the blue/yellow and grey violins in the second panel). The statistical comparison in the bottom panel shows that once the offsets are calculated, the observed luminosities for both post-mergers and pair galaxies are consistent within $1\sigma$ with those of their control ensembles. There is a diversity of $\Delta\mathrm{log(L_{X})}$ within the post-merger and pair samples, the extent of which is illustrated by the violin series in the third panel. Possible explanations for the moderate luminosities seen in the galaxy pair and post-merger AGN populations are explored further in Section~\ref{Discussion}. Later in Section~\ref{Multi-wavelength observability of X-ray AGNs in mergers}, we present evidence for the probable attenuation of X-ray fluxes in late stage mergers, and previous studies (e.g., \citealp{2021MNRAS.506.5935R,2021ApJS..257...61Y}) have shown evidence for increasing obscuration with decreasing galaxy separation. In this context, the luminosities of galaxy pairs with $r_{p}<20$ kpc and post-mergers might actually be intrinsically elevated, and attenuated by the obscuring material. However, a definitive test of any X-ray luminosity enhancement would require measurements of attenuation and intrinsic fluxes. 

\subsection{Multi-wavelength AGN excesses}
\label{Multi-wavelength AGN excesses}

Our X-ray results derived from eROSITA observations suggest that galaxy mergers are capable of triggering (or enhancing) and fueling AGNs. Even considering the effects of obscuration (and the connection between late-stage mergers and obscuration), X-rays from the engine in the centre of the AGN are less easily obscured than the BLR, and less conditional than NLAGN or mid-IR AGN detection (see \citealp{2016MNRAS.457..110M}). Still, a better physical understanding can be reached by supplementing our X-ray data with observations at other wavelengths.

\subsubsection{Individual AGN criterion excesses}
\label{Individual AGN criterion excesses}

\begin{figure*}
\includegraphics[width=\textwidth]{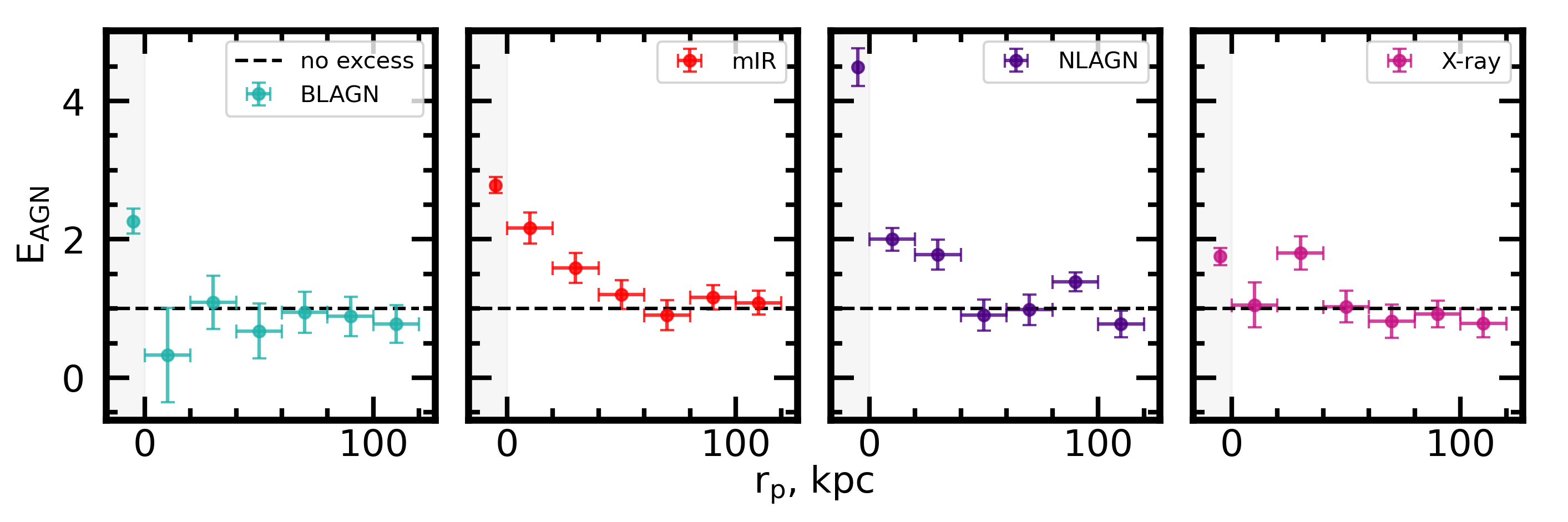}
\caption{Multi-wavelength excesses for BLAGNs, mid-IR, NLAGNs, and X-ray AGNs plotted side by side for comparison. BLAGN, mid-IR, and NLAGN statistics are not computed pairwise, since we have unblended measurements for each individual galaxy. The eRASS1 X-ray AGN excess is the same as in Figure~\ref{fig:xagn_exc}.}
\label{fig:4pan_excs}
\end{figure*}

Figure~\ref{fig:4pan_excs} offers an introduction to the multi-wavelength characteristics of the galaxy pairs and mergers in our sample. The first, second, and third panels show the AGN excess for interacting galaxy pairs and post-mergers in the style of Figure~\ref{fig:xagn_exc} for galaxies with a BLR detection, dust-obscured AGN detection, and NLR detection, respectively. The pair-phase results in these three panels use the entire sample of pair galaxies, rather than MGPs, since individual galaxies in a pair are always resolved in SDSS fibre spectroscopy and unWISE photometry. The full pairs sample is preferable where it can be reliably studied since it improves the statistics, but the qualitative results are in agreement when the MGPs are used instead. The eRASS1 excess series shown in the fourth panel is identical to that plotted in Figure~\ref{fig:xagn_exc}.

In the AGN unification model (\citealp{1993ARA&A..31..473A}; \citealp{1995PASP..107..803U}; and see \citealp{2018ARA&A..56..625H} for a recent review), the observability of the BLR is considered a function of viewing angle, and when the BLR is not visible it is due to obstruction by a dusty torus. The lack of a strong positive or negative signal for galaxy pairs in the first panel of Figure~\ref{fig:4pan_excs} suggests that either 1) pair-phase interactions do not preferentially trigger AGNs with accretion rates strong enough to change the covering fraction or depth of obscuring material in the nucleus (\citealp{2008MNRAS.385L..43F}; \citealp{2017Natur.549..488R}), or 2) pair-phase interactions are triggering and obscuring BLAGNs with similar frequencies, giving the appearance of no effect. After coalescence, however, BLRs are observed in excess of what is seen in isolated controls, indicating that the conditions brought on by coalescence facilitate observability of the BLR. One interpretation of the emergence of a BLAGN excess in post-mergers is that a subset of SMBHs in post-mergers reach high enough accretion states to remove obscuring material.

The excess of AGNs in the mid-IR series in the second panel of Figure~\ref{fig:4pan_excs} increases steadily from right to left, from wide pairs with 40<$\mathrm{r_{p}}$<60 kpc down to the post-merger epoch. The increasing trend of mid-IR AGN incidence with $\mathrm{r_{p}}$ is consistent with the underlying hypothesis of this work: that AGNs are triggered with increasing frequency as merger events progress. The signal is also affected by the increasing likelihood of obscuration on either galactic or nuclear scales over the course of the merger sequence (\citealp{2018MNRAS.478.3056B}). It is therefore unsurprising that this panel shows the smoothest evolution as a function of $\mathrm{r_{p}}$.

The NLAGN trend (third panel of Figure~\ref{fig:4pan_excs}) also suggests an increasing likelihood of AGN triggering with decreasing $\mathrm{r_{p}}$. The conditions for the excitation and detection of the NLR seem to be particularly common in post-mergers, resulting in the highest excess for any plot in this work at more than a factor of four. The particular NLAGN signal recovered is sensitive to the choices of S/N and BPT diagram criteria used, but the trend is not. We experimented with different AGN selection criteria to investigate the robustness of our NLAGN result. The excess signal is slightly weaker (with a difference of about 0.2-0.4 in the bins with a positive excess) when galaxies above the \citet{2003MNRAS.346.1055K} criterion on the BPT diagram are included along with those exceeding the \citet{2001ApJ...556..121K} maximum starburst criterion, suggesting that major mergers are preferentially associated with galaxies whose emission lines are unambiguously NLR-dominated. Lowering the S/N criterion for the BPT diagram emission lines (e.g., to S/N$>3$) has no statistically meaningful influence on the signal shown in the NLAGN panel of Figure~\ref{fig:4pan_excs}. We conclude that the quantitative NLAGN excess results are only weakly sensitive to the choice of NLAGN criteria, and our interpretation of the results is even less sensitive.

Taken together, the multi-wavelength AGN excesses shown in Figure~\ref{fig:4pan_excs} suggest more frequent AGN triggering and obscuration by gas clouds and dust as mergers progress during the pair phase, with circum-nuclear obscuration often clearing via radiation pressure (e.g., \citealp{2008MNRAS.385L..43F}) in the post-merger epoch. Within the merger sequence, AGNs are most likely to be triggered in the post-merger epoch (as suggested by the mid-IR and NLAGN panels, which are less impeded by the effects of obscuring material). Notably, post-mergers appear to be unique, either in their ability to trigger AGNs with accretion rates strong enough to remove obscuring material from the nucleus, or in that enough time has passed in a high accretion state that obscuring material has been gradually removed.

\subsubsection{Any AGN criterion}
\label{Any AGN criterion}

\begin{figure}
\includegraphics[width=\columnwidth]{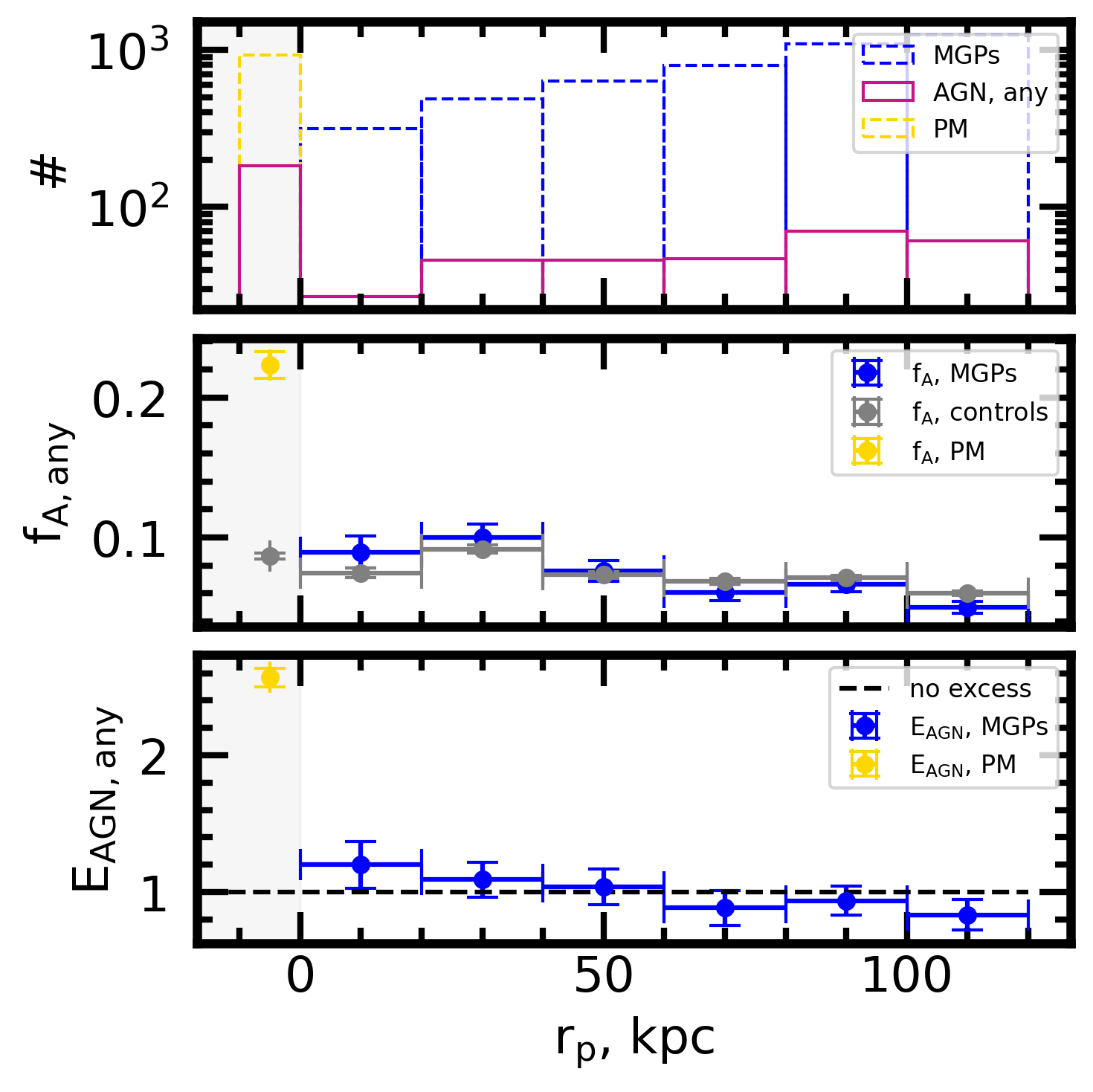}
\caption{“Any” AGN excess for the same pair and merger samples as Figure~\ref{fig:xagn_exc}. We find an excess of 1.2 in close galaxy pairs, and a dramatic 2.6$\times$ excess in post-mergers when we allow for any AGN criterion to be counted as a detection.}
\label{fig:anyagn_exc}
\end{figure}

None of our AGN selections can be truly complete, with NLAGNs and dust-obscured AGNs requiring the presence and detection of the excited / heated obscuring material, BLAGN detection requiring an exposed nucleus, and X-ray detection requiring a relatively high accretion rate (due to our relatively shallow threshold of $\mathrm{L_{X}}$\textasciitilde$10^{40}$ erg/s) and lack of heavy obscuration. Together, a more complete AGN sample can be assembled, although there are certainly gaps in the selection (e.g., where narrow-line sources probably host AGNs but are counted as "composite" sources between the \citealp{2001ApJ...556..121K} and \citealp{2003MNRAS.346.1055K} criteria on the BPT diagram, where a BLR is detected weakly with S/N<5 in the \citealp{2019ApJS..243...21L} catalogue, or when an AGN is not luminous enough in X-rays to reach the eRASS1 flux limit). Still, we can approach completeness by counting galaxies meeting any of our criteria as an AGN, and compute an excess.

We plot the results of an any-AGN excess study in Figure~\ref{fig:anyagn_exc}. For this experiment, we again must consider MGPs instead of individual pair galaxies, since X-ray detections are counted. If either pair galaxy is detected as a BLAGN, mid-IR, NLAGN, or X-ray AGN, we count the MGP as an AGN. We use the same criteria to classify pairs of matched control galaxies (the same matched controls as in Figure~\ref{fig:xagn_exc}) and apply the correction derived in Section~\ref{Pairwise treatment of X-ray detections in galaxy pairs}. Post-mergers are considered individually and matched to batches of single controls, so no correction is required.

The AGN fractions in this experiment and in Figure~\ref{fig:anyagn_exc} are likely the most accurate presented in this work: $6-10$\% for isolated galaxies depending on $\mathrm{M_{\star}}$ and $z$, $8-10$\% for close galaxy pairs, and as much as 22\% for post-mergers. These fractions are accurate insofar as the pair and post-merger samples selected are representative of their respective classes, which in turn depends on the selection described in Sections~\ref{Pair sample}, ~\ref{Post-merger sample}, and ~\ref{Control pool}. We note that our post-merger multi-wavelength AGN fraction is substantially lower than that reported in \citet{2023ApJ...944..168L} (77\%). The difference could be attributed to the fact that \citet{2023ApJ...944..168L} use a more inclusive S/N criterion on the BPT diagram for their optical selection, and their X-ray AGN sample is dominated by galaxies with lower $\mathrm{L_{X}}$, with the X-ray AGN threshold set at a $2-10$ keV luminosity of \textasciitilde$10^{40.5}$ erg/s.

We find small and statistically insignificant excesses of 1.04 and 1.09 for galaxy pairs between 40<$\mathrm{r_{p}}$<60 kpc and 20<$\mathrm{r_{p}}$<40 kpc, respectively. For galaxy pairs, the any-AGN approach favours control galaxies over the pairs, with many controls that were counted as non-AGNs in our single-criterion results (in the denominator of $f_{A}$ in the excess panels of Figure~\ref{fig:4pan_excs}) receiving an AGN classification when we include more AGN types. The reclassification of some control galaxies acts to diminish the AGN excess seen in pairs compared with Figure~\ref{fig:4pan_excs}. The loss of a significant excess in galaxy pairs with $\mathrm{r_{p}}$>20 kpc is a mathematical consequence of the higher baseline AGN fraction amongst galaxies in general when any AGN criterion is allowed (note the higher control AGN fractions in this figure compared to Figure~\ref{fig:xagn_exc}). The higher global AGN fraction proportionally shrinks the difference in AGN fraction between the pair and control samples, which constitutes the excess signal. Nonetheless, for close galaxy pairs with $\mathrm{r_{p}}$<20 kpc, we recover a positive and >1$\sigma$ excess of 1.2. Post-mergers with any AGN classification remain highly in excess of isolated controls with the same, by a factor of 2.6.

\subsubsection{Obscuration excess}
\label{Obscuration excess}

\begin{figure}
\includegraphics[width=\columnwidth]{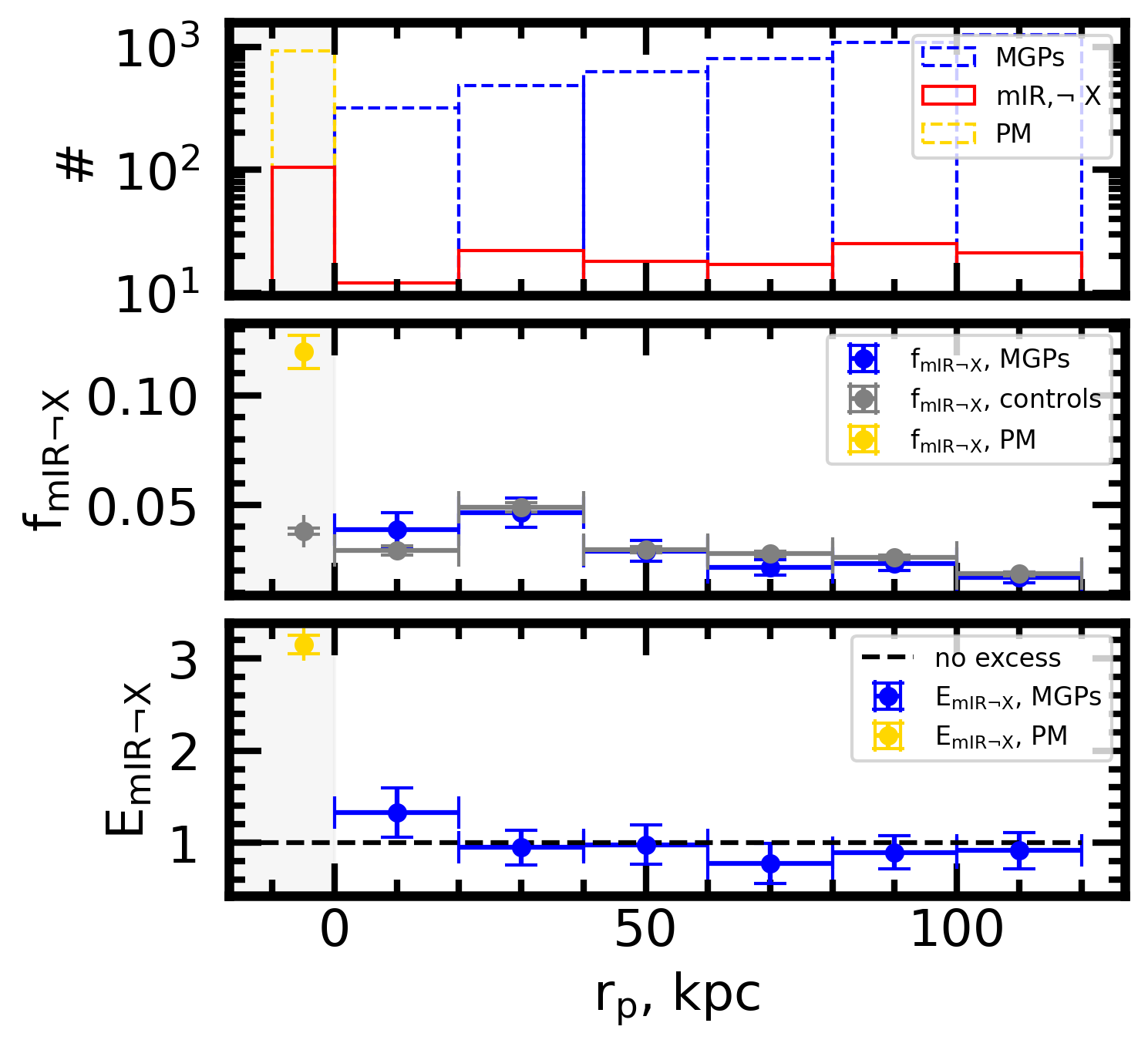}
\caption{Investigating the heavily obscured AGN class in close pairs and post-mergers. The top panel shows number of pairs where neither member has an X-ray detection, and one or both members have W1-W2<0.5. For post-mergers, the mid-IR and X-ray measurements for the target galaxy are assessed individually. The middle shows the fraction meeting the obscured AGN criterion, no X-ray plus W1-W2>0.5, in the merger and control samples. The bottom panel shows the excess. We note that AGNs in the numerator of the fraction plotted in the second panel likely includes galaxies that are only thinly obscured but intrinsically faint in the X-ray, in addition to heavily obscured AGNs. By matching controls on $\mathrm{M_{\star}}$ and $z$, we control for secular factors and luminosity distance so that the excess represents mid-IR AGNs whose X-ray non-detection is a consequence of the merger event.}
\label{fig:ctagn_exc}
\end{figure}

One of the main caveats of our X-ray results is that unconstrained degrees of obscuration are affecting the AGN fractions in eRASS1 for late-stage galaxy pairs and post-mergers. Obscuration effects have been implied by trends shown up to this point in Figures~\ref{fig:xagn_exc} and~\ref{fig:4pan_excs}, and strengthened by the assumption that many heavily obscured and dusty AGNs should be recoverable as power-law dust-obscured galaxies in the mid-IR. We take this a step further and investigate particularly the incidence of galaxies that are detected as AGNs in the mid-IR but not in the X-ray.

Figure~\ref{fig:ctagn_exc} has identical construction to Figure~\ref{fig:anyagn_exc}, except we count galaxies in the numerator of $\mathrm{f_{mIR\neg X}}$ when MGPs, post-mergers, or controls have W1-W2>0.5 in the mid-IR, and are not detected in eRASS1. For MGPs and their control pairs, we require eRASS1 non-detections for both members, and W1-W2>0.5 for at least one member. Post-mergers are again counted individually. It is worth noting that galaxies meeting the $\mathrm{mIR\neg X}$ criterion are not necessarily Compton-thick AGNs: some X-ray AGNs may have only thin obscuration, and are too faint to be detected in eRASS1. The middle panel, $\mathrm{f_{mIR\neg X}}$, should therefore be interpreted not as a Compton-thick fraction, but as a diverse subset that includes both heavily obscured galaxies and intrinsically X-ray faint galaxies in some proportion. Our control-matching methodology should remove any signal to do with luminosity distance, $\mathrm{M_{\star}}$, or secular processes. The bottom panel therefore represents the excess of AGNs that are observable in the mid-IR and unobservable in the X-ray specifically due to a merger.

The lower panel of Figure~\ref{fig:ctagn_exc} shows that there is no excess of mid-IR detected, X-ray undetected AGNs down to $\mathrm{r_{p}}=20$ kpc. Below 20 kpc, however, mergers are associated with $\mathrm{mIR\neg X}$ AGNs 1.31$\pm$0.27 times as often than isolated galaxies. Since the excess is designed to trace the influence of the merger, it is reasonable to assume that these are heavily obscured (or Compton thick) cases owed to a merger-induced influx of dusty gas (\citealp{2018MNRAS.478.3056B}). Such cases are likely the cause of the unity X-ray AGN excess measurement for MGPs with $\mathrm{r_{p}}$<20 kpc in Figure~\ref{fig:xagn_exc}.

Somewhat surprisingly, even though post-mergers exhibit a significant excess of X-ray AGNs compared to controls in Figure~\ref{fig:xagn_exc}, we find here that they also host (presumably heavily obscured or Compton-thick) $\mathrm{mIR\neg X}$ AGNs 3.1 times more often than matched controls. This large $\mathrm{mIR\neg X}$ excess suggests a diversity of observability in post-mergers, where heavy columns of obscuring material are sometimes removed by radiation pressure from the AGN, and sometimes continue to enshroud the nucleus. If we assume that the proportion of $\mathrm{f_{mIR\neg X}}$ AGNs in post-mergers in excess of $\mathrm{f_{mIR\neg X}}$ in controls are indeed heavily obscured cases, it would suggest that the $f_{A}$ measurement for X-ray luminous post-mergers is underestimated by \textasciitilde8\%. In turn, we can compute an "obscuration-corrected" X-ray AGN excess:

\begin{equation}
	\mathrm{E_{X,corr}=\frac{f_{X,PM}+(f_{mIR\neg X,PM}-f_{mIR\neg X,Ctrl})}{f_{X,Ctrl}}}
	\label{dustcorr}
\end{equation}

Applying this correction to post-mergers, we calculate a new excess of \textasciitilde4.1, bringing the measurement into agreement with our NLAGN result in Figure~\ref{fig:4pan_excs}. Even though NLAGNs and X-ray AGNs only overlap partially, and NLAGNs do not trace well the high-$\mathrm{L_{X}}$ tail of the X-ray AGN sequence, the degree of overlap between NLAGN detections and X-ray AGN detections appears to remain relatively consistent as a function of merger stage. As such, AGNs selected using the NLAGN criteria used in this work mirror the trends (but not the total fraction) of X-ray AGN incidence in galaxy mergers.

\subsection{Multi-wavelength observability of X-ray AGNs in mergers}
\label{Multi-wavelength observability of X-ray AGNs in mergers}

There are several interesting multi-wavelength intersections to explore in the context of the merger sequence. We next inspect the overlap between different AGN diagnostics as a function of merger stage.

\begin{figure*}
\includegraphics[width=\textwidth]{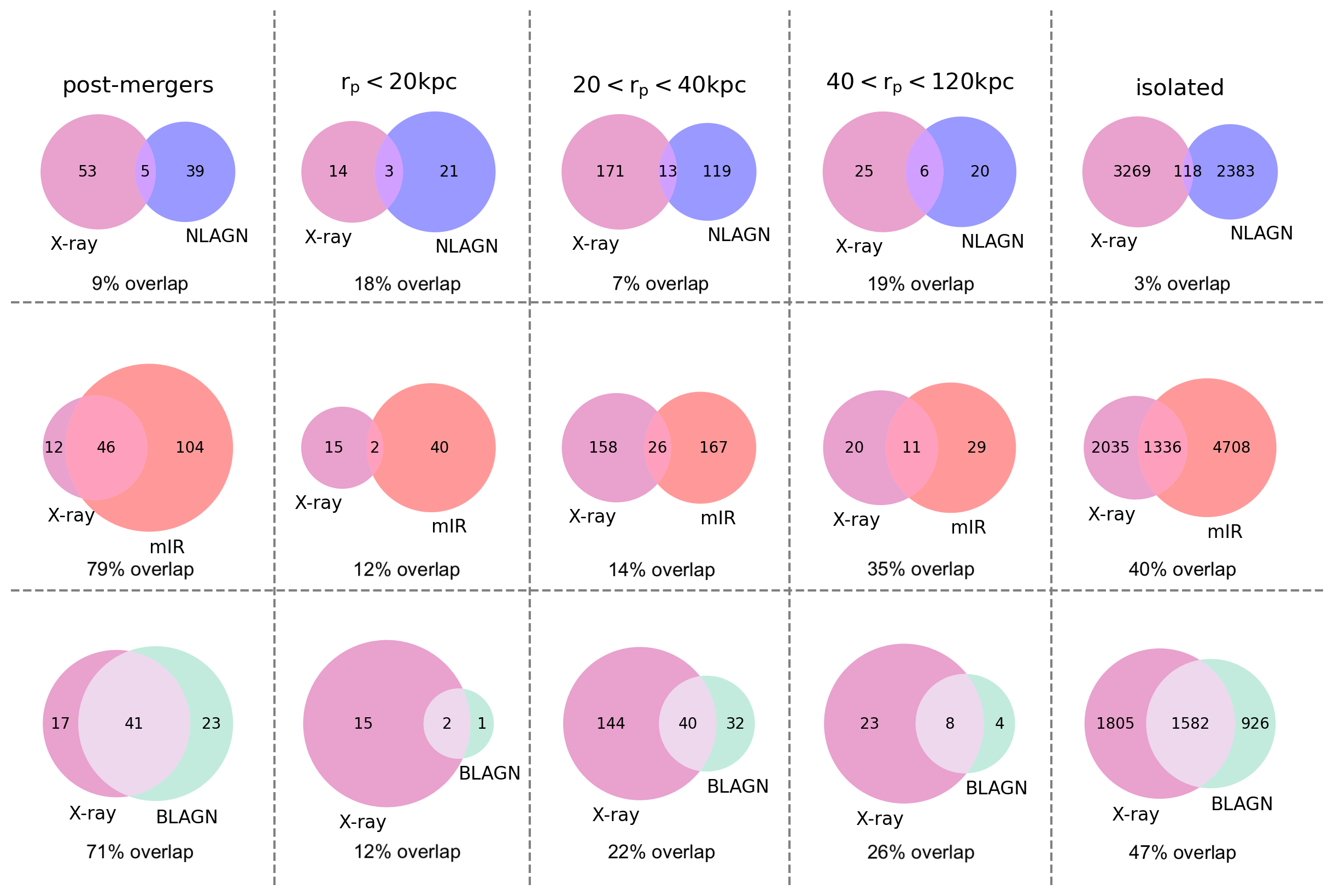}
\caption{The multi-wavelength observability of X-ray AGNs in coarse bins of projected separation. We have taken the isolated (control pool) galaxies, galaxy pairs where we see little to no signal on our excess plot ($40-120$kpc), and the two bins of interacting galaxy pairs where we uncover significant merger-AGN connections in this work (<20, and $20-40$ kpc), as well as post mergers. The percentage of X-ray AGNs co-detected as an AGN via the second diagnostic is stated in each panel as well.}
\label{fig:multi_venn}
\end{figure*}

Figure~\ref{fig:multi_venn} shows the multi-wavelength characteristics of post-mergers, three subsets of pair galaxies, and isolated galaxies. The trends of the dual-detected (X-ray plus another criterion) samples with merger stage reflect and verify the trends presented earlier (e.g., in Figure~\ref{fig:4pan_excs}). The lowest two $\mathrm{r_{p}}$ bins span the same $\mathrm{r_{p}}$ windows as those in the Figures (e.g., Figure~\ref{fig:xagn_exc}) above, and a third bin is added to characterize generally any wider pairs with $\mathrm{r_{p}}$>40 kpc, for which we do not uncover a strong merger-AGN connection in this work. These five merger sequence categories are plotted in columns. In rows, we plot the overlap between the eRASS1 X-ray AGN sample and the AGN sample recovered using our NLAGN criterion (first row), mid-IR (second row), and BLAGN (third row) using Venn diagrams. The width of each Venn diagram is normalized, so that the horizontal overlap between adjacent diagrams can be compared directly. The percentage of X-ray AGNs that are co-detected as an AGN using the second diagnostic are also listed below the Venn diagrams in each panel.

The result for NLAGNs mirrors what we find in Section~\ref{Individual AGN criterion excesses} -- the overlap between narrow-line and X-ray detected AGNs is small, but comparatively stable as a function of merger stage. NLR observability is therefore less influenced by the turmoil of an ongoing or completed galaxy merger, even though the fraction of the X-ray AGN sample recovered using the narrow-line criteria is relatively low.

The mid-IR and BLAGN results tell the same story from two sides. Amongst isolated galaxies, 47\% of X-ray AGNs are co-detected as BLAGNs, and mid-IR AGNs detection coincides with X-ray AGN detection in 40\% of isolated X-ray cases. The BLAGN - X-ray and mid-IR - X-ray overlap fractions in isolated galaxies are likely decided by the orientation of the torus and accretion rate (see Figure~\ref{fig:agn_fracs}). As mergers progress, X-ray and mid-IR AGNs increasingly separate from one another as dust-obscured AGNs are more often heavily obscured. At the same time, galaxies with a detected BLR become increasingly rare on account of the same physics.

In post-mergers, we find much stronger overlap between the eRASS1 X-ray, mid-IR, and BLAGN samples, since the accretion rates brought on by many mergers are sufficient to blow out circum-nuclear dust and expose the BLR. A meaningful subset of the mid-IR post-mergers that are not seen in the X-ray, meanwhile, are presumably either hopelessly obscured or lacking the strength to push the obscuring material away as explored in Figure~\ref{fig:ctagn_exc}.

\begin{figure*}
\centering
\includegraphics[width=15cm,height=15cm,keepaspectratio]{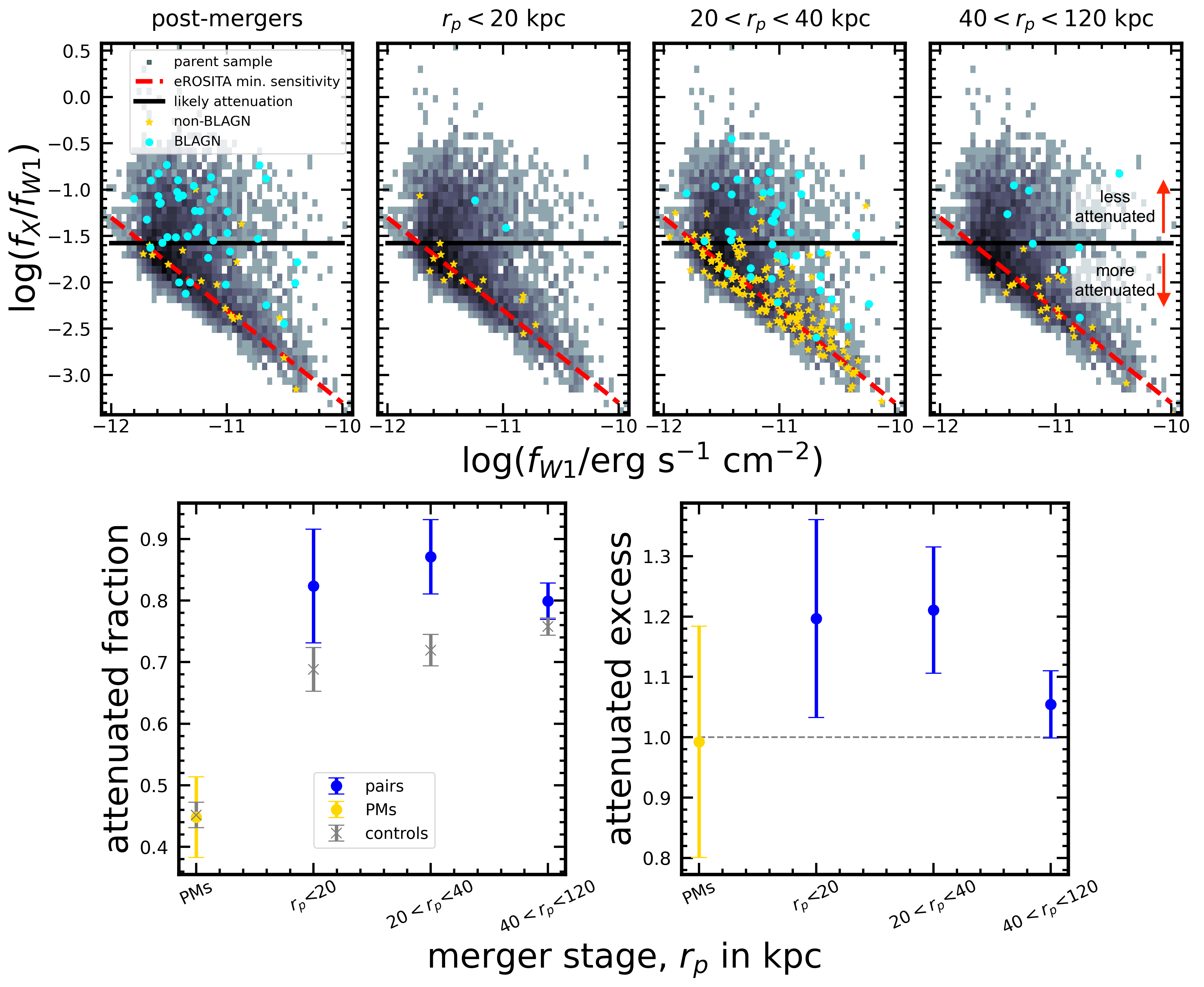}
\caption{Another view of the evolution of AGN observability as a function of merger stage, using the same four bins of galaxy mergers as in Figure~\ref{fig:multi_venn}. For the top four panels, post-mergers are shown in the leftmost panel, galaxy pairs with $r_{p}<20$ kpc are shown in the second, pairs with $20<r_{p}<40$ kpc are shown in the third, and pairs with $40<r_{p}<120$ kpc are shown in the fourth. All panels show the ratio of observed fluxes in the X-ray (as observed by eROSITA) and in the W1 mid-IR band (from unWISE) as a function of the W1 flux. The figure shows the number of mergers belonging to each category that have the multi-wavelength appearance of being partially obscured or unobscured as a function of merger stage. In each panel, we plot the flux ratio corresponding to the minimum observable flux density for eROSITA using the red dashed line. In the bottom left panel, we plot the fraction of merger-AGNs appearing below the black line (in the region associated with X-ray attenuation by obscuring material) in each of the four panels above, with errors computed as the binomial error on the fraction, $\sqrt{f(1-f)/N}$. Post-mergers are shown in yellow, and galaxy pairs belonging to each of the main categories from the panels above are shown in blue. We also plot the fraction of attenuated galaxies in isolated control samples matched to the mergers on $\mathrm{M_{\star}}$ and $z$. In the bottom right panel, we plot the attenuation excess for the merger samples over the controls.}
\label{fig:fluxrat_comp}
\end{figure*}

The physical scenario suggested by Figures~\ref{fig:ctagn_exc} and~\ref{fig:multi_venn} is explored further in Figure~\ref{fig:fluxrat_comp}, which investigates the ratio of fluxes in the X-ray and mid-IR (unWISE W1) as a function of the W1 flux. Flux in W1 is converted to erg/s/cm$^{2}$/\AA~ from the values reported in the unWISE data release (\citealp{Lang_2016}), and multiplied by the effective width of the W1 band, which spans \textasciitilde$2.8-3.8$ $\mu$m. For eROSITA, we use the total maximum likelihood X-ray flux from eRASS1. Since galaxies included in Figure~\ref{fig:fluxrat_comp} must have an X-ray flux, they are all counted as AGNs in this study. Galaxies with unobscured AGNs are expected to have fairly consistent X-ray to mid-IR flux ratios (the upper ``cloud'' population in Figure~\ref{fig:fluxrat_comp}), especially in hard X-rays (e.g., \citealp{2009ApJ...693..447F,2012ApJ...754...45I,2019MNRAS.484..196T}). Cases where the X-ray/W1 flux ratio is lower are likely the result of degrees of attenuation, an effect that is particularly relevant due to the relatively soft X-ray regime to which eROSITA is sensitive (lower ``sequence'', Figure~\ref{fig:fluxrat_comp}).

Unlike Figure~\ref{fig:ctagn_exc}, which investigates the frequency with which obscuration results in the loss of X-ray AGNs from the sample, Figure~\ref{fig:fluxrat_comp} shows how often X-rays are attenuated (but still detected) as a function of merger stage. To verify the connection between the amount of obscuration and the position of AGNs on the diagram in Figure~\ref{fig:fluxrat_comp}, we colour-code the data points for post-mergers and pairs based on whether they were (teal data series) or were not (yellow series) detected as BLAGNs. Unsurprisingly, BLAGNs are commonly found in the region of the diagram associated with little obscuration, while other merger-AGNs are generally found in the lower sequence of X-ray attenuated AGNs. The flux ratio corresponding to the minimum sensitivity of eROSITA for a given W1 flux is marked with the red dashed line. AGNs whose X-rays are obscured to the point of non-detection by eROSITA would populate the vacant region below and to the left of the red dashed line. We also draw an empirical separation between the unobscured and attenuated populations at log$(f_{X}/f_{W1})=-1.575$ (black line, Figure~\ref{fig:fluxrat_comp}). The distribution of log$(f_{X}/f_{W1})$ has two peaks (corresponding to the mode values for the unattenuated and attenuated populations), and $-1.575$ lies between them. In the bottom panel of Figure~\ref{fig:fluxrat_comp}, we also compute the fraction of mergers appearing below the empirical attenuation criterion in each of the four merger stage bins. We also make a statistical comparison between the mergers and matched samples of non-interacting control galaxies, following the same methodology described in Section~\ref{Control pool}. The attenuated fractions for the matched controls are shown in the grey data series in the bottom left panel of Figure~\ref{fig:fluxrat_comp}. In the bottom right panel, we compute the attenuation excess by taking the ratio of the attenuated fractions in the merger and matched control samples.

Our earlier results have shown that galaxy pairs with projected separations $40<r_{p}<120$ kpc do not host AGNs in excess of controls (see Figure~\ref{fig:4pan_excs}), leading to only a few galaxy pairs plotted in teal (for BLAGNs) and red (for non-BLAGNs) in the rightmost top panel of Figure~\ref{fig:fluxrat_comp}. Even at wide separations, we detect an elevated fraction of sources with X-ray attenuation, suggesting that inflows may already be common in early-stage (but still post-pericentre) galaxy pairs. Closer merging pairs with $20<r_{p}<40$ kpc do host elevated incidence rates of NLAGNs, mid-IR AGNs, and X-ray AGNs (Figure~\ref{fig:4pan_excs}), and the corresponding panel in Figure~\ref{fig:fluxrat_comp} is thus more populated. The majority of galaxy pairs in the third panel of Figure~\ref{fig:fluxrat_comp} are X-ray attenuated and non-BLAGNs, suggesting that merger-induced gas inflows are both triggering and obscuring AGNs in galaxy pairs with separations of $20<r_{p}<40$ kpc. The second panel in Figure~\ref{fig:fluxrat_comp} suggests that X-ray detected galaxy pairs with  $r_{p}<20$ kpc are very often attenuated, and rarely host BLAGNs. Meanwhile, Figure~\ref{fig:ctagn_exc} indicates that many AGNs become so obscured at close separations that they are no longer detected in eROSITA. The loss of X-ray AGNs likely accounts for the rarity of galaxy pairs in the second panel of Figure~\ref{fig:fluxrat_comp} relative to the adjacent panels, as well as the lack of an X-ray AGN excess for pairs with $r_{p}<20$ kpc. The leftmost panel of Figure~\ref{fig:fluxrat_comp} shows that a large proportion of post-mergers appear in the unobscured ``cloud'', consistent with our earlier finding in Figure~\ref{fig:4pan_excs} that they frequently host visible BLAGNs. We also find that circum-nuclear blowout is not ubiquitous after coalescence, since a number of post-mergers remain attenuated, and some lack a BLR detection. A diversity of intrinsic accretion rates and degrees of obscuration in post-mergers are likely responsible for this result, as well as the distribution of observed X-ray luminosities in post-mergers presented earlier in Figure~\ref{fig:lum_enh}. A temporal hypothesis could also explain the diversity of attenuation seen in post-mergers, since more recent post-mergers would have had less time to clear obscuring material.

\section{Discussion}
\label{Discussion}

The results of this work are best collated chronologically, starting with the first infall of a galaxy that will eventually merge, and ending after coalescence when the merging galaxies' nuclei become indistinguishable from one another. In general, our findings suggest that it is essential to keep track of two AGN phenomena: SMBH triggering / accretion, and obscuration.

Even without the influence of a merger, isolated galaxies studied in this work appear to span the entire dynamic range of AGN properties. In general, they have relatively low AGN fractions of about $6-10$\% (between $25-50$\% as many as in late-stage galaxy mergers, see Figure~\ref{fig:anyagn_exc}). When AGNs are present, they appear to have (on average, and perhaps by definition) moderate luminosities governed by the strength of the secular accretion events that have triggered them. Their observability depends mainly on this luminosity, and is modulated by the orientation of the BLR, dusty torus, and NLR relative to the line of sight (see Figures~\ref{fig:agn_fracs} and~\ref{fig:multi_venn}).

After galaxies have experienced their first pericentric passage in a major merger (here approximated as 0.1<$\mu$<10, since our selection targets such galaxies) the tidal torques begin to funnel gas towards the galaxies' centres and onto their SMBHs. We find that this effect is visible for pairs with $\mathrm{r_{p}}$<40 kpc. While small projected separations do not guarantee post-pericentre status, simulations suggest that pairs at these separations are advanced in the merger sequence: \citet{2024MNRAS.529.1493P} recently reconstructed the orbits of pre-merger galaxy pairs from IllustrisTNG 100-1, and \textasciitilde91\% of galaxies in the \citet{2024MNRAS.529.1493P} sample with physical separations between $20<r<40$ kpc, $\Delta v<300$ km/s, and $0.1<\mu<10$ have experienced at least one pericentric passage with their interacting companion. Inflows ushered in by close pair-phase interactions begin triggering AGNs with higher frequency than in isolated galaxies, with an excess of \textasciitilde$1.5-2$ depending on the AGN flavour studied. While AGNs are more common, they are observed to have statistically indistinguishable uncorrected X-ray luminosities relative to AGNs in isolated galaxies (Figure~\ref{fig:lum_enh}), and partial attenuation by obscuring gas likely plays a role in determining the observed X-ray luminosities of many well-separated galaxy pairs (third panel, Figure~\ref{fig:fluxrat_comp}). In well-separated galaxy pairs, we do not find evidence for elevated incidence of heavy obscuration leading to removal from the eROSITA catalogue ($20<r_{p}<40$ kpc, Figure~\ref{fig:ctagn_exc}). While pair-phase interactions are therefore better at triggering detectable AGNs than secular processes, the relative strengths of secular and merger-induced accretion are not well constrained in eROSITA due to obscuration effects.

AGNs are still being triggered and fueled at the closest separations (Figures~\ref{fig:4pan_excs} and~\ref{fig:anyagn_exc}), and their uncorrected X-ray luminosities are still typical relative to AGNs in isolated galaxies. In the closest galaxy pairs with $\mathrm{r_{p}}$<20, as with Figure~\ref{fig:xagn_exc}, it is likely that degrees of obscuration (but not total obscuration, as the galaxies studied in Figure~\ref{fig:lum_enh} have eRASS1 detections and are not Compton-thick) of the X-ray source also contribute to the appearance of a relatively sparse (when obscuration leads to an eROSITA non-detection) and average-luminosity (when obscuration attenuates the detected flux) X-ray AGN population. We find that an increasing number of AGNs are obscured by dense (potentially Compton-thick, although we do not have measurements of $N_{H}$ from eROSITA) dust columns in late stage pairs (lowest $\mathrm{r_{p}}$ bin, Figure~\ref{fig:ctagn_exc}). This worsens the reliability of AGN identification methods relying on observations from the SMBH engine, and likely attenuates the X-ray luminosities of galaxies that are detected (Figure~\ref{fig:fluxrat_comp}). The degree of attenuation is not explored in this work, but future observations allowing characterization of the dust column in heavily dust-obscured AGNs will allow this connection to be explored with greater nuance.

When coalescence occurs, there is a noticeable increase in the frequency of AGNs. Depending on the AGN selection criteria, AGNs are \textasciitilde$2-4$ times more common in post-mergers compared to isolated controls (Figure~\ref{fig:4pan_excs}). Due to the effects of heavy obscuration in late stage mergers, we argue that these quantitative results do not fully represent the intrinsic accretion states of the SMBHs. In a subset of cases, we expect based on models (e.g., \citealp{2008MNRAS.385L..43F}) and observations (e.g., \citealp{2017Natur.549..488R}) that coalescence-epoch gas inflows are fueling AGNs that remain heavily obscured by dense obscuring gas columns (\citealp{2018MNRAS.478.3056B}). Even more than in close pairs, however, obscuration bears on the observability of AGNs in post-mergers (Figure~\ref{fig:ctagn_exc}), so the true AGN excess is likely \textasciitilde4.1. For the same reasons as in close galaxy pairs, we expect that the true unobscured luminosity enhancements of the post-mergers are somewhat higher than what is computed in Figure~\ref{fig:lum_enh} (see the left panel of Figure~\ref{fig:fluxrat_comp}), but more detailed characterization of the obscuring column would require X-ray spectroscopy (e.g., an analysis similar to that in \citealp{2023ApJ...954..116P}). Since post-mergers best represent the cumulative effect of a merger event on galaxy evolution, deep X-ray spectroscopy of post-merger samples in the future would help to develop a more nuanced understanding. The ubiquity, demographics, and observability of dual SMBHs in late stage galaxy mergers are lately being explored in multi-wavelength studies (e.g., \citealp{2017MNRAS.470L..49E}; \citealp{2021ApJ...911..100T}; \citealp{2023ApJ...942L..24K}), and future case studies on the topic will lend additional context to the statistical results presented in this work. Broadly, the results of this work serve as priors on AGN observability in mergers and isolated galaxies that should be carefully accounted for in subsequent studies of AGN incidence or characteristics.

In comparing our results with the literature, we restrict ourselves to a discussion of the low-redshift Universe. In a sample of 43 post-mergers from the \citet{2013MNRAS.435.3627E} sample and a matched sample of 430 non-interacting control galaxies, \citet{2020MNRAS.499.2380S} check for $3\sigma$ hard ($2-10$ keV) X-ray detections from XMM Newton and find an excess of $2.22^{+4.44}_{-2.22}$. In this work, we calculate an excess of $1.8\pm0.1$. We posit that the smaller post-merger sample and lower statistical power in the \citet{2020MNRAS.499.2380S} experiment is primarily responsible for the difference in their results compared to ours, since the excess quantities computed are in good agreement. Additional differences may follow from the X-ray AGN selection used in \citet{2020MNRAS.499.2380S}, since eROSITA is sensitive to softer X-rays than XMM Newton. Still, the error bars on the AGN fractions and excess in \citet{2020MNRAS.499.2380S} wholly encompass the equivalent measurements in this work, so any discrepancies owed to AGN selection methodology are secondary to the influence of sample size.

The \citet{2023ApJ...944..168L} post-merger sample is selected by eye from SDSS and Legacy Survey (which includes DECaLS) imaging, and the AGNs in \citet{2023ApJ...944..168L} are identified using $2-10$ keV fluxes from both Chandra and XMM (for the X-ray), BPT diagram criteria similar to this work (for optical AGNs; though, with a minimum emission line S/N of $>3$, while we use S/N$>5$) and the same WISE W1$-$W2>0.5 colour criterion as this work (in the mid-IR). Compared to matched controls,  \citet{2023ApJ...944..168L} find post-merger AGN excesses of $2.4-2.6$, $3.4-3.8$, and 6.7 for X-ray, NLAGNs, and mid-IR AGNs, respectively. In this work, we find AGN excesses of 1.8, 4.5, and 2.8 for the nearest experimental analogues for X-ray, NLAGNs, and mid-IR AGNs. \citet{2023ApJ...944..168L} also use radio observations from the VLA FIRST survey to check for an excess of radio cores in post-mergers, but do not find a significant excess. The multi-wavelength analysis of \citet{2023ApJ...944..168L} is therefore in qualitative agreement with our own, with both studies indicating a multi-wavelength AGN excess of \textasciitilde$2-4$ depending on the criteria used. The quantitative differences between our results and those of \citet{2023ApJ...944..168L} could be caused in part by the fact that the multi-wavelength AGN fraction reported in \citet{2023ApJ...944..168L} is much higher than our own (77\%), owing to differences in multi-wavelength AGN selection (particularly for X-ray AGNs, since \citealp{2023ApJ...944..168L} capture less-luminous X-ray AGNs). We also combine our study of heavily obscured AGN candidates (in Figure~\ref{fig:ctagn_exc}) with our X-ray AGN results (Figure~\ref{fig:xagn_exc}) and attempt a correction for the effects of obscuration. Implementing the correction, we estimate that the intrinsic (or obscuration-corrected) X-ray AGN excess in post-mergers is \textasciitilde$4.1$, also in statistical agreement with the result presented for AGNs with $\mathrm{L_{X}}>10^{42}$ or $\mathrm{L_{X}}>10^{43}$ in Figure 11 of \citet{2023ApJ...944..168L}.

Our suggestion of a high obscured AGN fraction in close galaxy pairs is in tension with \citet{2023ApJ...943...50H}, who find that Compton-thick obscuration is not responsible for the low incidence rate of AGNs in a sample of 7 galaxy pairs (14 galaxies) interacting within $r_{p}<5$ kpc. We instead posit that in late-stage galaxy pairs, AGN feedback has already begun to expel circum-nuclear gas, leading in turn to lower accretion rates. These two effects are not easily distinguished using the observations in this work. We therefore allow for the possibility that the $\mathrm{mIR\neg X}$ AGN population studied in Figure~\ref{fig:ctagn_exc} could contain both AGNs that are obscured by dusty gas brought in by merger events and AGNs whose accretion rates are diminished by feedback. The late-stage pair feedback scenario is not obviously compatible the multi-wavelength analysis presented in Section~\ref{Multi-wavelength AGN excesses}, since both NLAGNs and mid-IR AGNs are still being triggered in late galaxy pairs, but the timescales for dust cooling (for the mid-IR case; \citealp{2017ApJ...844...21I}) and narrow-line recombination (for the NLAGNs; \citealp{2013ApJ...779..109P}) could allow for sustained detection of narrow-line and mid-IR AGNs even after AGN feedback has diminished the instantaneous accretion rate. Conversely, the detection of a merger excess amongst luminous and obscured AGNs in \citet{2018Natur.563..214K} suggests that obscuration and accretion can coincide in late stage mergers. \citet{2018MNRAS.478.3056B} uncover an analogous result in simulations, reporting that mergers achieve simultaneous maxima in both $\mathrm{N_{H}}$ and $\mathrm{L_{AGN}}$ at the time of coalescence (Figure 1 of \citealp{2018MNRAS.478.3056B}). It is therefore likely that some proportion of our $\mathrm{mIR\neg X}$ AGNs are in fact heavily obscured. Other studies tracing AGNs via hard X-rays in close galaxy pairs (e.g., \citealp{2012ApJ...746L..22K,2016ApJ...825...85K}) have found a merger-AGN connection; such results are consistent with the hypothesis that the relatively soft X-ray response of eROSITA may be responsible for a number of X-ray undetected AGNs in this work. Since BLAGNs are typically only observable through low column densities ($N_{H}<10^{22}$/cm$^{2}$), the co-detection of many X-ray AGNs as BLAGNs in this work indicates that our X-ray AGN sample is generally not heavily obscured (\citealp{2022ApJS..261....4O}). Our suggestion that obscuration can often clear post-coalescence is in qualitative tension with the findings of \citet{2022MNRAS.514.4828M}, who find no statistical evidence for an elevated incidence rate or luminosity of ionized outflows in post-mergers. The tension can potentially be reconciled by the fact that \citet{2022MNRAS.514.4828M} require MPA-JHU stellar masses for their work, and BLAGNs are nominally excluded from the MPA-JHU sample. In this work, meanwhile, an abundance of BLAGNs in post-mergers are the primary piece of evidence for preferential clearing of obscuration in the post-merger phase. Evidence for outflows could potentially be found using the method of \citet{2022MNRAS.514.4828M} in galaxies that are excluded from the \citet{2022MNRAS.514.4828M} post-merger sample, but a different stellar mass estimate and an adjusted spectral fitting procedure accounting for a BLR component would need to be implemented.

\section{Summary}
\label{Summary}

In this paper, we use analytical spectroscopic techniques to identify a sample of 4,565 merging galaxy pairs, and a hybrid (machine learning plus human classification) approach to identify a new sample of 923 post-mergers, all in the survey overlap between SDSS DR7, the first data release from the eROSITA all-sky survey, DECaLS $r$-band imaging, and unWISE forced mid-IR photometry. Using eROSITA to characterize the X-ray AGNs in the sample, plus a broader multi-wavelength study of the AGNs' characteristics, we find the following:

\begin{itemize}
\item High-X-ray-luminosity AGNs with observed (uncorrected for attenuation) X-ray fluxes $\mathrm{L_{X}}$>$10^{43}$ erg/s are very likely to be co-identified as AGNs in the mid-IR and as BLAGNs. As a result, matching X-ray samples (such as eRASS1) to optical samples that omit galaxies with broad lines (such as MPA-JHU) will miss many bona-fide X-ray AGNs. We suggest that cirum-nuclear dust is likely forced out via accretion disk radiation pressure in this luminosity regime (Figure~\ref{fig:agn_fracs}), giving rise to preferential observability of the BLR. Relatively few X-ray AGNs are also NLAGNs, and galaxies co-detected with X-ray and NLAGNs lie at lower (uncorrected) observed $\mathrm{L_{X}}$. We conclude that NLAGNs are therefore either intrinsically under-luminous, or more heavily obscured, leading to X-ray attenuation or removal from the X-ray sample.
\medskip
\item X-ray AGNs are more likely to be triggered in close galaxy pairs between $20-40$ kpc and post-mergers, with measured excesses \textasciitilde1.8 compared with a control sample of non-interacting galaxies. For very close galaxy pairs with projected separations under 20 kpc, we measure no excess of X-ray AGNs relative to controls. We suggest this result, and the specific value of the observed X-ray AGN excess in post-mergers, are affected by high rates of heavy obscuration in late stage mergers (Figure~\ref{fig:xagn_exc}).
\medskip
\item We find that the observed (not corrected for obscuration) X-ray luminosities in both post-mergers and interacting galaxy pairs are statistically comparable to what is seen in isolated galaxies. We also find that X-ray AGNs in galaxy pairs and post-mergers are shown to be preferentially attenuated (Figure~\ref{fig:fluxrat_comp}), and that X-ray AGNs in close pairs with $\mathrm{r_{p}}$<20 kpc and post-mergers appear to preferentially drop out of the eRASS1 sample due to obscuration (Figure~\ref{fig:ctagn_exc}). Combining these pieces of evidence, we argue that the statistically indistinguishable observed X-ray luminosities seen in galaxy mergers may result from elevated intrinsic X-ray luminosities in the merger sample and enhanced column densities in late stage mergers. Nevertheless, the $\mathrm{L_{X}}$ distributions in our merger samples indicate a wide diversity of SMBH accretion rates across the merger sequence.
\medskip
\item We present the results of a multi-wavelength AGN study that folds in mid-IR dust obscured, narrow-line, and broad-line optical AGN classifications. We find that BLAGNs are not preferentially triggered by pair-phase interactions (but do show a link to the post-merger phase), while mid-IR and NLAGNs are increasingly detected in close pairs and most often detected in post-mergers (Figure~\ref{fig:4pan_excs}).
\medskip
\item Considering all our AGN criteria simultaneously, we find a multi-wavelength AGN fraction of \textasciitilde22\% in our post-merger sample. We find that galaxy pairs with $\mathrm{r_{p}}$<20 kpc host AGNs in excess of isolated controls by a factor of \textasciitilde1.2, and post-mergers have a strong AGN excess of 2.6 (Figure~\ref{fig:anyagn_exc}).
\medskip
\item In an effort to identify potentially highly obscured or Compton-thick AGNs, we quantify the fraction of galaxies classified as AGNs in the mid-IR, but undetected in the X-ray. We find a 1.3$\times$ excess of such objects in galaxy pairs with $\mathrm{r_{p}}$<20 kpc, and a 3.1$\times$ excess in post-mergers. Galaxy pairs, then, are more likely to remain heavily obscured on account of their lower accretion rates. Meanwhile, post-mergers appear to have a chance of blowing out obscuring material in the nucleus (Figures~\ref{fig:xagn_exc},~\ref{fig:multi_venn}, and~\ref{fig:fluxrat_comp}).
\end{itemize}

Throughout this work, post-mergers are shown to be unique by several criteria: they more often host AGNs than secular or pair-phase galaxies, and they are a major channel for the production of both heavily obscured AGNs and type 1 AGNs with visible BLRs and accretion disks after the removal via radiation pressure of dusty gas in the nucleus.

\section*{Acknowledgements}
\label{Acknowledgements}

This work was completed on the unceded territory of the Lekwungen speaking Songhees, Esquimalt and $\mathrm{\ubar{\mathrm{W}}S\acute{A}NE\acute{C}}$ peoples, as well as that of the $\mathrm{S\ubar{\mathrm{k}}w\ubar{\mathrm{x}}w\acute{u}7mesh}$ speaking $\mathrm{x}^{\mathrm{w}} \mathrm{m} \mathrm{\rote} \uptheta \mathrm{k}^{\mathrm{w}} \mathrm{\rote} \mathrm{y} \mathrm{\rote} \mathrm{m}$ (Musqueam), $\mathrm{S\ubar{\mathrm{k}}w\ubar{\mathrm{x}}w\acute{u}7mesh}$ (Squamish), and s$\mathrm{\rote}$lilw$\mathrm{\rote}$ta\textipa{\textbeltl} (Tsleil-Waututh) Nations, who have stewarded the land for centuries and continue to do so today.

This work is based on data from eROSITA, the soft X-ray instrument aboard SRG, a joint Russian-German science mission supported by the Russian Space Agency (Roskosmos), in the interests of the Russian Academy of Sciences represented by its Space Research Institute (IKI), and the Deutsches Zentrum für Luft- und Raumfahrt (DLR). The SRG spacecraft was built by Lavochkin Association (NPOL) and its subcontractors, and is operated by NPOL with support from the Max Planck Institute for Extraterrestrial Physics (MPE).
The development and construction of the eROSITA X-ray instrument was led by MPE, with contributions from the Dr. Karl Remeis Observatory Bamberg \& ECAP (FAU Erlangen-Nuernberg), the University of Hamburg Observatory, the Leibniz Institute for Astrophysics Potsdam (AIP), and the Institute for Astronomy and Astrophysics of the University of Tübingen, with the support of DLR and the Max Planck Society. The Argelander Institute for Astronomy of the University of Bonn and the Ludwig Maximilians Universität Munich also participated in the science preparation for eROSITA.

We thank Prof. Jonathan Trump and Dr. Connor Bottrell for their helpful comments on this work with regard to physical models of AGNs and the availability of high-quality optical imaging data, respectively.

Data from the IllustrisTNG simulations are integral to this work. We thank the Illustris Collaboration for making these data available to the public.

This research was enabled, in part, by the computing resources provided by Compute Canada, and the Canadian Advanced Network for Astronomical Research (CANFAR).

\section*{Data Availability}
\label{Data Availability}

A catalogue of the post-mergers identified using our machine learning model and visual classifications is available as digital supplemental material along with the release of this paper. Additional data products can be made available upon request from the author. The unWISE (http://unwise.me/), MPA-JHU (wwwmpa.mpa-garching.mpg.de/SDSS/DR7/), \citet{2019ApJS..243...21L} BLAGN (cdsarc.cds.unistra.fr/viz-bin/cat/J/ApJS/243/21), and eRASS1 (www.mpe.mpg.de/eROSITA) catalogues are open data.



\bibliographystyle{mnras}
\bibliography{bib_01} 





\appendix

\section{Stellar mass fitting}
\label{A1}

The results in the paper are enabled by a new stellar mass catalogue produced using the SED fitting method of \citet{2016ApJS..227....2S} but using only SDSS $griz$ model magnitudes. In Section~\ref{Parent catalogue}, we refer to the fact that we exclude optical $u$ and UV bands from our stellar mass fitting routine, based on severe contamination of the UV bands by emission from the accretion disk in cases where the nucleus is unobscured (BLAGNs).

\begin{figure}
\includegraphics[width=\columnwidth]{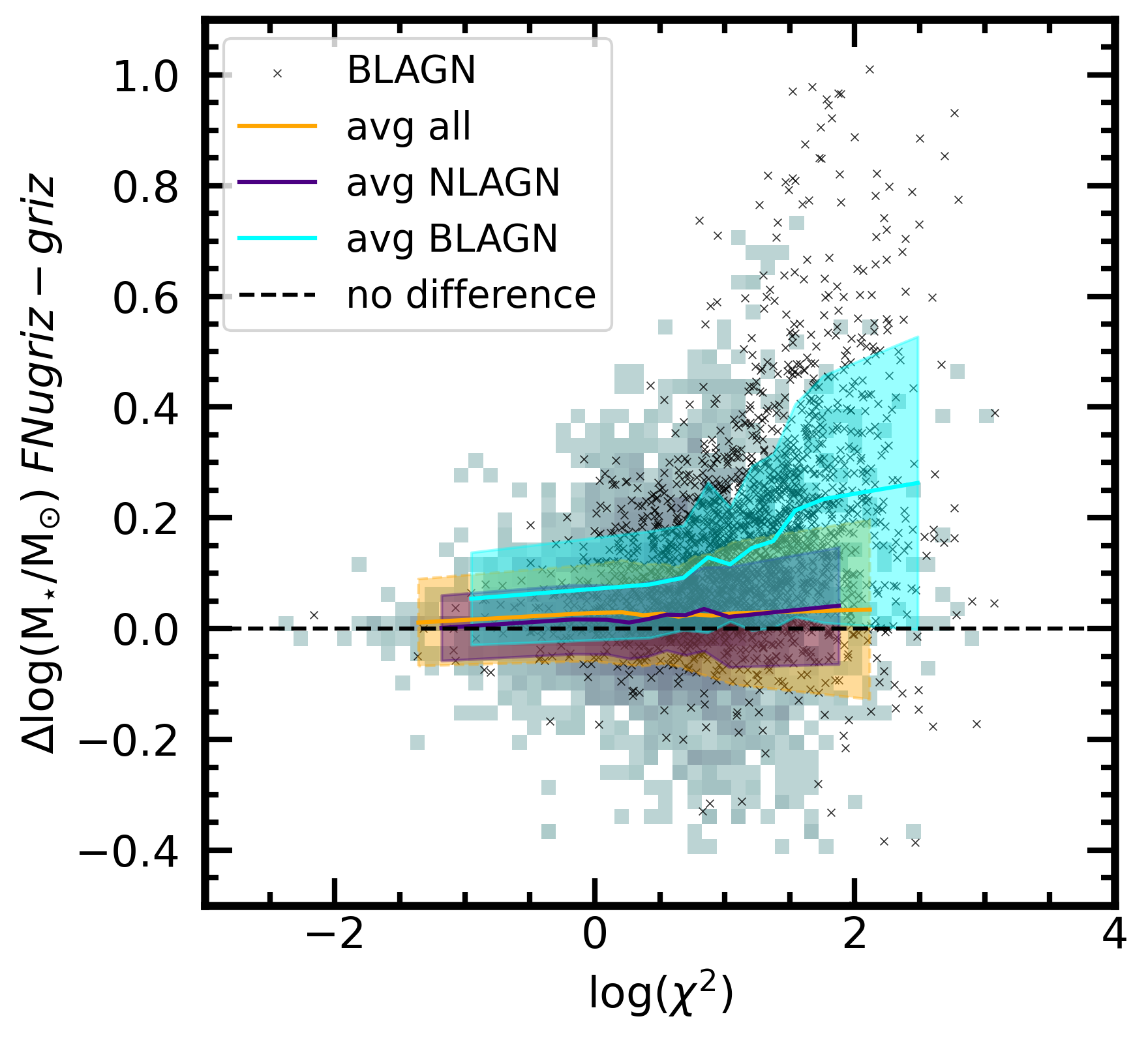}
\caption{A comparison of the stellar mass fitting results using $griz$ photometry from SDSS only, and GALEX far- and near-UV ($FN$) plus SDSS $ugriz$. We plot the logscale difference between the $FNugriz$ and $griz$ stellar masses against log(${\chi}^{2}$), representing the goodness-of-fit for the mass estimate. The 2D histogram in the background shows the extent of the data for a subset of parent catalogue galaxies appearing in GALEX that meet neither our NLAGN nor BLAGN criteria. The mean trend and 1$\sigma$ region for the non-AGN set are shown in orange. The indigo series and error region shows the mean trend for NLAGNs, and the black scatter points and cyan data series show the entire BLAGN sample appearing in GALEX, and the mean trend, respectively.}
\label{fig:massfit}
\end{figure}

In order to characterize the influence of the BLR and/or accretion disk on our $\mathrm{M_{\star}}$ fitting procedure (despite our choice to remove the bands most likely to be impacted), we make estimates of M$_\mathrm{\star}$ with and without $u$-band and UV (far and near UV, $F$ and $N$, respectively) photometry. We compare the two sets of M$_\mathrm{\star}$ estimates in Figure~\ref{fig:massfit}. On the horizontal axis, we plot log(${\chi}^{2}$), which represents the quality of the fit obtained through SED modelling that includes all of the bands ($FNugriz$). A high value of log(${\chi}^{2}$) indicates that the model developed for the stellar SED is not optimally suited to the observed SED of a galaxy, presumably because of AGN contamination. On the vertical axis, we plot the log difference between the stellar mass estimate obtained using $FNugriz$ and $griz$ photometry, such that positive values indicate an overestimate of the stellar mass when $FNu$ are included compared to using $griz$ only.

The background 2D histogram is for a subset of parent catalogue galaxies appearing in GALEX, and which are neither NLAGNs nor BLAGNs according to the criteria set in Section~\ref{AGN samples}. The mean trend and 1$\sigma$ error region for the non-NLR and non-BLR subset are shown in orange. The mean trend and error region for NLAGNs is shown in the indigo data series. Amongst NLAGNs, the mass offset is not highly sensitive to the quality of the fit (i.e., the indigo series is comparatively flat), although there is a small zero-point offset of \textasciitilde0.1 dex between the mass estimates made with and without $FNu$-band photometry. The offset is to be expected, since contribution to the total stellar mass by young, UV-bright stars is underestimated without UV fluxes included. Still, the lack of a trend amongst galaxies without an optical AGN and NLAGN hosts indicates that the quality of the fit does not lead to an over- or under-estimate of $\mathrm{M_{\star}}$ when galaxies fall within the parameters of the modelling assumptions.

For BLAGNs, however, we find that the predictions made including the $FNugriz$ bands are higher by as much as \textasciitilde1 dex (on average only \textasciitilde0.1 dex, since the statistics are dominated by broad-line AGNs with a weaker total contribution by the BLR to the SED) than those made with $FNu$ excluded. The trend turns sharply upward at high log(${\chi}^{2}$), corresponding to a population of problematic cases (note the vertical extent of the black scatter points). In these cases, the large fitting errors indicate that the model (which does not include BLAGN or accretion disk templates) is having difficulty reproducing the observed SED using stellar templates alone. In these cases, the $FNugriz$ masses are severely overestimated due to UV contamination by the unobscured SMBH accretion disk.

Further removal of the $g$ band does not significantly change the mass estimates for broad-line AGNs, so we conclude that excluding UV+$u$ photometry appears to correct for AGN contamination. Taken together, the trends shown in Figure~\ref{fig:massfit} indicate that $griz$ masses are optimal for our study, since they are not as affected by contamination from the AGN when it is exposed.


\bsp	
\label{lastpage}
\end{document}